\documentclass[10pt,a4paper]{article}

\usepackage{fullpage}
\usepackage{setspace}
\usepackage{parskip}
\usepackage{titlesec}
\usepackage{xcolor}
\usepackage{lineno}

\usepackage{times}

\usepackage{url}

\titlespacing{\section}{0pt}{*3}{*1}
\titlespacing{\subsection}{0pt}{*2}{*0.5}
\titlespacing{\subsubsection}{0pt}{*1.5}{0pt}

\usepackage{authblk}

\usepackage{graphicx}
\usepackage[space]{grffile}
\usepackage{latexsym}
\usepackage{textcomp}
\usepackage{longtable}
\usepackage{tabulary}
\usepackage{booktabs,array,multirow}
\usepackage{amsfonts,amsmath,amssymb}
\providecommand\citet{\cite}
\providecommand\citep{\cite}

\newif\iflatexml\latexmlfalse

\AtBeginDocument{\DeclareGraphicsExtensions{.pdf,.PDF,.eps,.EPS,.png,.PNG,.tif,.TIF,.jpg,.JPG,.jpeg,.JPEG}}

\usepackage[utf8]{inputenc}
\usepackage[ngerman,english]{babel}

\begin{document}

\title{\emph{Gender gaps in urban mobility~}}

\author[1,*]{Laetitia Gauvin}%
\author[1,*]{Michele Tizzoni}%
\author[1,2]{Simone Piaggesi}%
\author[3]{Andrew Young}%
\author[4]{Natalia Adler}%
\author[3]{Stefaan Verhulst}%
\author[5,6]{Leo Ferres}%
\author[1]{Ciro Cattuto}%
\affil[1]{~ISI Foundation, Torino, Italy}%
\affil[2]{~Doctoral School in Data Science and Computation, University of Bologna, Bologna, Italy}%
\affil[3]{~The Governance Lab, New York University, New York, NY, United States}%
\affil[4]{~United Nations International Children's Emergency Fund (UNICEF), New York, NY, United States}%
\affil[5]{~Data Science Institute, Universidad del Desarrollo, Santiago, Chile}%
\affil[6]{~Telefónica R\&D, Santiago, Chile}%
\affil[*]{~These authors contributed equally to this work.}%

\vspace{-1em}

  \date{}

\begingroup
\let\center\flushleft
\let\endcenter\endflushleft
\maketitle
\endgroup

\onehalfspacing

\newpage

\textbf{Abstract}

The use of public transportation or simply moving about in streets are gendered issues. Women and girls often engage in multi-purpose, multi-stop trips in order to do household chores, work, and study ('trip chaining'). Women-headed households are often more prominent in urban settings and they tend to work more in low-paid/informal jobs than men, with limited access to transportation subsidies.
Here we present recent results on urban mobility from a gendered perspective by uniquely combining a wide range of datasets, including commercial sources of telecom and open data. We explored urban mobility of women and men in the greater metropolitan area of Santiago, Chile, by analyzing the mobility traces extracted from the Call Detail Records (CDRs) of a large cohort of anonymized mobile phone users over a period of 3 months. We find that, taking into account the differences in users' calling behaviors, women move less than men, visiting less unique locations and distributing their time less equally among such locations. By mapping gender differences in mobility over the 52 \textit{comunas} of Santiago, we find a higher mobility gap to be correlated with socio-economic indicators, such as a lower average income, and with the lack of public and private transportation options. Such results provide new insights for policymakers to design more gender inclusive transportation plans in the city of Santiago.

\section{Introduction}

Cities, and how they are designed, are not gender-neutral. Consider daily mobility: simply moving around provides for different experiences depending on whether you are a woman or a man. Insecurity and the fear of physical or sexual violence in public spaces 
and when using public transportation are key factors that limit the everyday movement of women and girls~\cite{loukaitou2014fear}. 
The location of bus stops, or how well-lit streets are, can also greatly affect womens' movement. Women and girls also engage more in multi-purpose, multi-stop trips ('trip chaining') in order to do household chores, as well as other gender differentiated roles~\cite{brown2014urban}. Women-headed households are often more prominent in urban settings and they tend to work more in low-paid or informal jobs than men, with limited access to transportation subsidies~\cite{tacoli2012urbanization}. 

At the same time, urbanization offers many possibilities to reduce gender gaps through a wealth of new opportunities. However, urbanization also increases inequalities by, for example, reinforcing geographical segregation, especially in developing-world settings~\cite{chant2013cities}. Mobility is a critical factor in reducing such segregation. Investigating the role of gender in urban mobility is key to better understanding of whether women and young girls can fully benefit from opportunities offered by cities, and in the process realize their human rights.
 
While several cities around the globe are starting to pay more attention to women's experiences, the unique mobility needs of women and girls are rarely taken into account in urban and transportation planning. 
One reason involves entrenched gendered power hierarchies, present in most societies~\cite{uteng2016gendered}. Yet another reason involves the absence of robust data about the lives of women and girls - especially as it relates to daily mobility. As pointed out by the World Bank and by initiatives such as Data2X, studies on pressing issues such as fighting poverty and hunger or epidemics, among many others, have suffered from a lack of gender disaggregated data, effectively assuming that the problems of different genders are equivalent~\cite{weltbank_gender_2011,buvinic2014mapping}. Collapsing gender disparities prevents relevant organizations and policy-makers from getting a full picture of reality and ultimately limits the possibility of intervening to bridge the gendered gap that does exist. Non-existent and substandard data on gender differences have many drawbacks and effectively means that urban planning is often gender-blind~\cite{buvinic2016closing}. If there is no understanding on how women travel around differently, certain transportation options that could support women and girls may be overlooked. 

Most traditional mobility studies are derived from surveys based on relatively few observations over a limited time span, or with low spatial resolution or other errors (e.g., due to self-reporting)~\cite{groves2006nonresponse, couper2000web, clarke1981error}. More importantly, the question of how observed differences in mobility can be explained by innate sex-related differences, such as physical differences, or by gendered socially constructed factors, such as household roles, remains highly debated~\cite{rosenbloom2004understanding}. 
Also, long-term trends of gender differences in mobility are intertwined with global demographic and socioeconomic trends and are therefore hard to capture. This is particularly true in urban areas, where the population is continuously and rapidly growing; according to the United Nations,  the world's urban population is projected to account for almost 70\% of total global  population by 2050~\cite{nations2018}. 

Mobility is a complex issue and no single dataset or approach is sufficient to unpack its multidimensionality and offer insights on the way forward for decision-makers. In addition, the Data2X report on Big Data and the Well Being of Girls points out that much of the data that could provide new insights on these issues is collected by corporations, and is therefore often not available to researchers and public policymakers~\citep{Vaitla2017}.

Nowadays, with the pervasiveness of mobile devices, it has become possible to achieve large-scale urban sensing and explore the mobility of individuals at unprecedented scale~\cite{Blondel_2015,Naboulsi_2016}. 
In the past 10 years, many studies have successfully used Call Details Records (CDRs) to extract and analyze human mobility patterns~\cite{Gonz_lez_2009,Song_2010,Calabrese_2013,Beir__2018,Graells_Garrido_2017}. 
However, even if gender differences in mobility patterns derived from mobile phone data have been sometimes investigated, they have been mostly considered as somewhat peripheral~\cite{Song_2010} or comparatively small-scale~\cite{Psylla_2017}.

In this work, we study urban mobility from a gendered perspective in the greater metropolitan area of Santiago, Chile. With almost 7 million inhabitants, Santiago's metro is one of the largest metropolitan areas of South America, and like many cities across the continent, Santiago continues to expand and sprawl~\cite{lovera2015urban}. 
Its population already accounts for about 40\% of the national population and it is projected to grow steadily in the next decades~\cite{puertas2014assessing} posing many challenges to urban planners and policy makers, especially related to the design and adaptation of new and existing transportation infrastructures to serve the complex mobility needs of its inhabitants. 

Our study had two main objectives. First, to assess and quantify gender disparities in the mobility patterns of Santiago residents;  and, second, to identify socio-demographic factors and the availability of transport options that are associated with mobility inequalities. 
To achieve these objectives, we analyzed the mobility traces extracted from Call Detail Records of a large cohort of anonymized mobile phone users disaggregated by sex (male or female) over a period of 3 months. We then mapped indicators of mobility differences between males and females to 51 \emph{comunas} (Spanish for municipalities) of the Santiago Metropolitan Region and we investigated the association between mobility inequalities and socio-demographic indicators in different areas of the city, as well as their relationship with the Santiago transportation network structure. Finally, we added a ``semantic layer" to the mobility patterns of Santiago residents by identifying specific points of interests that are more frequently present along women's or men's trajectories in the urban space, thus demonstrating how our approach can identify specific gendered mobility needs.

\section{Results}

\subsection{Gender inequalities in mobility}

We analyze the mobility patterns of 418,624 individuals extracted from about 2 billion anonymized Call Detail Records (CDRs) collected between May 1 and July 30, 2016.
Our anonymized users' sample carries information about users' sex and socioeconomic status (see Materials and Methods for definitions). 
The sample is highly representative of the socio-demographic structure of the Santiago Metropolitan Region in terms of population, gender ratio and socioeconomic group distributions at the level of \textit{comuna} (see Fig.~S1 in the Supplementary Materials).

We assess gender differences in mobility by computing four mobility metrics for each individual and by evaluating the gender effect size through estimation statistics~\cite{ho2018moving}.
We characterize the mobility behavior of Santiago residents by looking at: i) the number of distinct locations visited by a user during the study period, $N_l$, ii) the number of distinct locations that account for at least 80\% of a user calling activity, $\hat N_l$, iii) the Shannon mobility entropy, $S$, and iv) the radius of gyration, $r_{g}$. 
To distinguish locations, the study area has been divided into 726 cells of about 1 km$^2$ regularly spaced according to the position of the cell towers as explained in the Materials and Methods. 
In the following, the words \textit{location} and \textit{cell} are used interchangeably.

\begin{figure}[th!]
\begin{center}
\includegraphics[width=1.00\columnwidth]{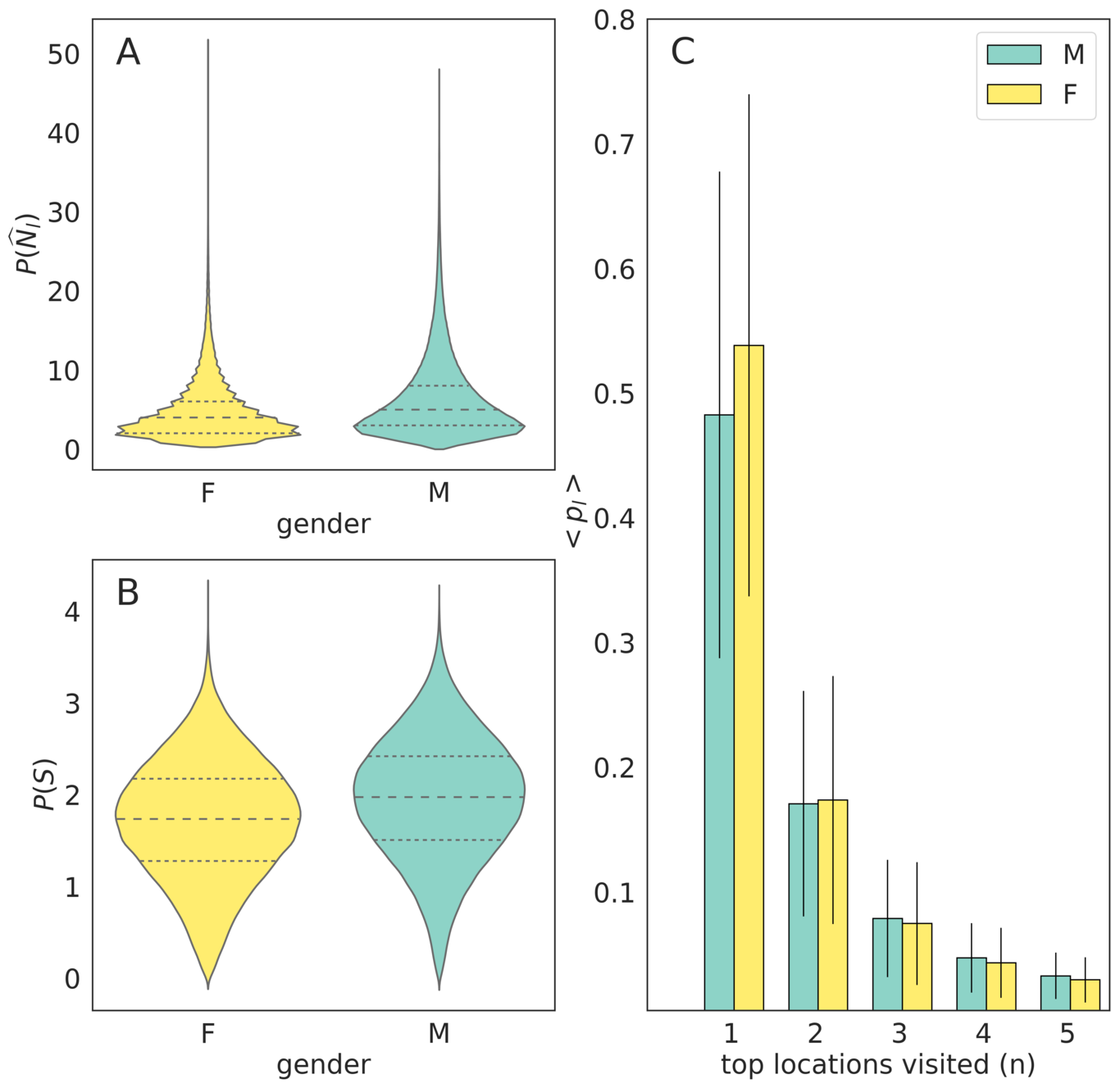}
\caption{{\textbf{Distributions of mobility metrics by gender.} Violin plots show
the distributions by gender of the number of locations accounting for
80\% of a user's activity (\textbf{A}) and the users' Shannon mobility
entropy (\textbf{B}). Women visit fewer locations and their movements
are characterized by a smaller entropy. Panel~\textbf{C} shows the
distributions of the mean probability of visiting the 5 most frequented
locations of each user, by gender. Error bars correspond to the standard
deviation of the mean.
{\label{413292}}%
}}
\end{center}
\end{figure}

We first look at gender differences in the total number of unique locations visited by the users. We observe that women travel to fewer unique locations than men.
Specifically, considering the complete set of locations visited by a user, we find that - over 3 months - women have traveled about nine locations less than men, on average: $\Delta N_l = \langle N_l \rangle_{M} - \langle N_l \rangle_{F} = 8.58$, $95\%$ CI $[8.43, 8.73]$.
If we only look at unique locations that belong to the core of users' daily activity, $\hat{N}_l$, the difference between men and women becomes $\Delta \hat{N}_l = 2.02$, $95\%$ CI $[1.98, 2.05]$ as shown in Fig.~\ref{413292}A.

Another key aspect that characterizes human mobility is the average distance traveled by an individual. In our sample, the average radius of gyration of women, $\langle r_{G} \rangle_F $, is $1.09$ Km ($95\%$ CI $[1.07, 1.12]$) shorter than the average men's radius of gyration, $\langle r_{G} \rangle_M$. 
Thus, women movements tend to be more spatially localized than men, as shown by the distributions of $r_{g}$ by gender (see Fig.~S2 in the Supplementary Materials).

We then look at the diversity of mobility patterns, measured by the Shannon entropy. 
Fig.~\ref{413292}B shows that women movements are consistently characterized by a smaller entropy compared to men, $\Delta S = 0.26$ ($95\%$ CI $[0.26, 0.27]$), indicating that women distribute their trips among a few highly preferred locations, while men distribute their trips among many locations with almost equal probability.
Accordingly, we observe that women can be more frequently found at their most visited location. Indeed the frequency rank plot of all visited locations for all users (Fig.~{\ref{413292}}C) shows a higher frequency of visits, $\langle p_i \rangle$, for women's first and second ranked locations, while the women to men ratio of $\langle p_i \rangle$ reverses from the third ranked location onward. 

\selectlanguage{english}

Since we infer mobility patterns from users' calling activity, it is natural to ask whether the observed disparities are due to gender differences in mobile phone usage. 
Indeed, in our sample, women call less frequently than men, for an average of 4.73 calls/day compared to 5.21 calls/day.
We answer this question in two ways (see Text A of Supplementary Materials for more details). 
First, we restrict our analysis to users' with the highest calling activity, i.e. those who made at least 3 calls per day on average over the 3 month period, thus increasing the number of observations for men and women in our dataset. 
Focusing on the most active users, of which 51\% are women, the gender differences in mobility become larger than those observed with the original user sample, with a smaller CI ($\Delta \hat{N}_l = 2.73$, $95\%$ CI $[2.68, 2.79]$).
Second, we down-sample the call activity of men by removing up to 50\% of the calls made by men in our original dataset and recomputing the Shannon entropy of all users. Even after doing so, women's entropy remains significantly smaller than men's.
We also check the robustness of our findings when varying the time frame of our analysis.
Overall, all users' mobility metrics, entropy and number of visited locations, display a high stability (Pearson correlation $r = 0.81$, $p<10^{-3}$), when measured over time windows of 9 days and considering each interval separately (see Text A of the Supplementary Materials).
Also, women and men distribute their calls through the day in the same way, displaying a very similar activity pattern by hour (see Fig.~S3 in the Supplementary Materials), hence we can rule out that the observed gender differences in entropy are trivially associated with gender differences in temporal activity patterns.

\subsection{Gender mobility and socio-demographic indicators}

\begin{figure}[th!]
\begin{center}
\includegraphics[width=1.00\columnwidth]{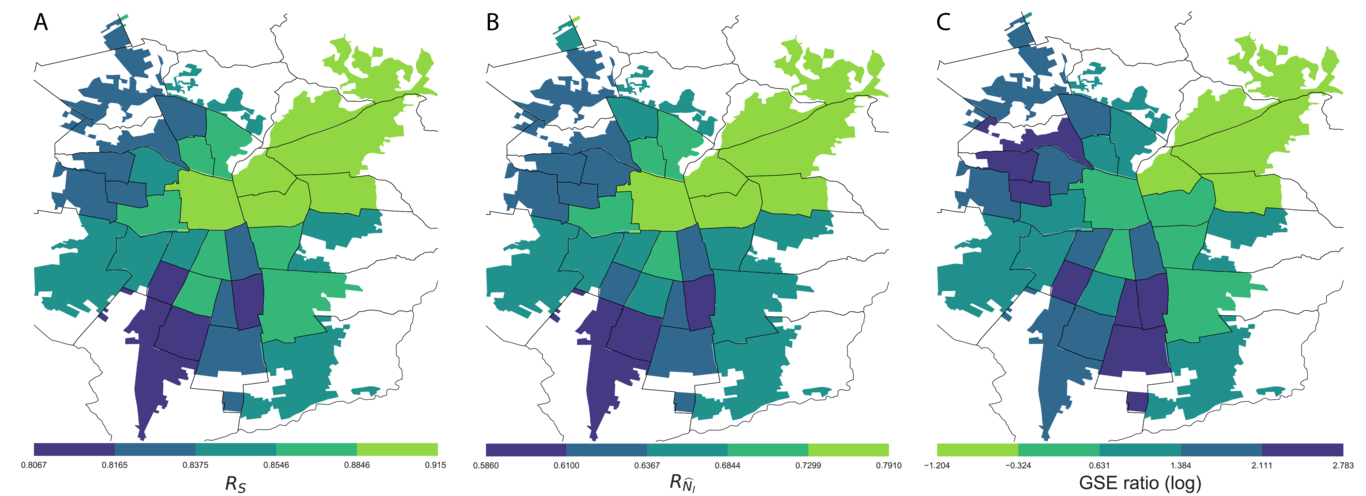}
\caption{{\textbf{Spatial patterns of gender mobility inequalities and
wealth.~}Choropleth maps of the metropolitan area of Santiago showing
the women to men ratio of entropy (\textbf{A}) and of the~number of
locations accounting for 80\% of users' activity~(\textbf{B}) by
\emph{comuna}. Panel~\textbf{C} shows the spatial distribution of
the~GSE ratio. Black lines indicate the administrative boundaries of
the~\emph{comunas.~}The boundary of the colored area corresponds to the
urban area of Santiago as defined by the National Institute of Statistics (INE).
{\label{750203}}%
}}
\end{center}
\end{figure}\selectlanguage{english}

To better understand what are the factors underlying the observed gender disparities in mobility, we examine their relationship with a number of socio-demographic indicators.

First, we examine the relationship between the socioeconomic status of the users and their mobility metrics. 
To this aim, we evaluate the gender differences in mobility through estimation statistics after splitting the users into 5 socioeconomic groups (``Grupos Socio-Econ\'omicos" or GSE, in the Spanish acronym) that are defined by income and household status (see Materials and Methods for the exact definition of GSE and Fig.~S4 in the Supplementary Materials for the distribution of users by GSE and gender). 
Table \ref{Tablediff} shows that the gender gap in all the mobility metrics widens as the socioeconomic status of the users decreases (ABC1 being the wealthiest segment and E being the most deprived).
For instance, $\Delta \hat{N}_l$ grows from 1.66 for users in the ABC1 group to 2.50 for those in the group E. In other words, poorer women tend to remain more localized than their male counterparts. 
Interestingly, although smaller, the gender gap in mobility is observed even among users who belong to the wealthiest class, thus indicating that a full mobility equality is not achieved even in presence of a high income. 

To further investigate how gender differences in mobility vary across different socioeconomic strata of the population, we perform a spatial analysis of the gender mobility gap by first assigning a home location to each user, based on their call activity (see Materials and Methods for details). 
In this way, we are able to examine gender mobility disparities at two resolutions: a very fine ``grid" resolution given by the cell locations and a coarser resolution corresponding to the \textit{comunas} of the Santiago Metropolitan Region.

We measure the gender mobility gap in a given location or \textit{comuna} by computing the women to men ratio of two mobility metrics, $S$ and $\hat{N}_{l}$, averaged over all users who live in that location (see Materials and Methods).
We denote the women to men ratios as $R_S$ and $R_{\hat{N}_{l}}$, respectively.
Panels \textbf{A} and \textbf{B} of Fig.~\ref{750203} show choropleth maps of the Santiago Metropolitan Area displaying the spatial variation of $R_S$ and $R_{\hat{N}_{l}}$ across 30 \textit{comunas} of the urban Santiago.
The map in panel \textbf{C} of Fig.~\ref{750203} shows the spatial distribution of wealth by \textit{comuna}, measured by the GSE ratio, that we define as the ratio between the number of residents belonging to the socioeconomic groups C3, D, and E, and the number residents belonging to a higher GSE (see  Materials and Methods for the exact definition of GSE ratio).
In general, Santiago displays a high level of segregation. 
The wealthiest \textit{comunas}, characterized by a lower GSE ratio are mostly located in northwestern Santiago, while areas in the eastern and the southern outskirts are home to the poorest residents and display a high GSE ratio. 
A similar segregation pattern is also evident in the spatial distribution of the gender mobility gap: the gap increases significantly when moving from the wealthiest to the most deprived \textit{comunas} of Santiago.
In particular, we measure the semi-partial Pearson correlation coefficient~\cite{fisher1924distribution} between $R_S$, $R_{\hat{N}_{l}}$ and the GSE ratio, by controlling for the variations in the call activity by gender and the differences in the sex ratio across \textit{comunas}. 
Both $R_S$ and $R_{\hat{N}_{l}}$ are strongly and negatively correlated with the GSE ratio, with correlation coefficients $r=-0.59$ ($p<0.001$) and $r=-0.53$, ($p<0.001$), respectively. 
Thus as the GSE ratio decreases, in the wealthiest areas of Santiago, both $R_S$ and $R_{\hat{N}_{l}}$ converge to the unit value, corresponding to gender equality in mobility (see Fig.~S5 in the Supplementary Materials for the scatterplot of $R_S$ and $R_{\hat{N}_{l}}$ against the GSE ratio).   

We further investigate the relationship between the sociodemographic characteristics of the Santiago Metropolitan Region and the gender differences in mobility, by considering a number of census variables as predictors of the gender gap (see Materials and Methods for the complete definitions of the variables). 
Table \ref{TableCorr} reports the semi-partial Pearson correlation coefficients computed between different sociodemographic variables and both $R_S$ and $R_{\hat{N}_{l}}$, in 51 \textit{comunas} of the SMR, for which we have at least 1,000 users.

The gender gap in mobility is significantly correlated with the gender gap in employment, suggesting that employment status may explain the observed differences in mobility behavior.
On the other hand, gender differences in education levels are not significantly associated with the gender gap in mobility.
Childcare duties are often considered a significant cause of mobility inequalities for women.  
Indeed, we find that a higher fertility rate is associated with a larger gender mobility gap.
If we look at the household structures in different \textit{comunas}, we find that a higher presence of large households (i.e. including dependent relatives or children) tends to be associated with a higher inequality between women and men in terms of mobility, with women staying more put, probably to bear the brunt of childcare duties. 
Conversely, in those \textit{comunas} where a larger proportion of households is formed by a single person (either a woman or a man), mobility patterns of men and women are more similar.  
But instead, a higher proportion of single parents living with their children -- which can be reasonably thought of being mostly single mothers -- is associated to a larger gender difference in mobility patterns. 

\subsection{Gender mobility and access to transport}

\begin{figure}[th!]
\begin{center}
\includegraphics[width=1.00\columnwidth]{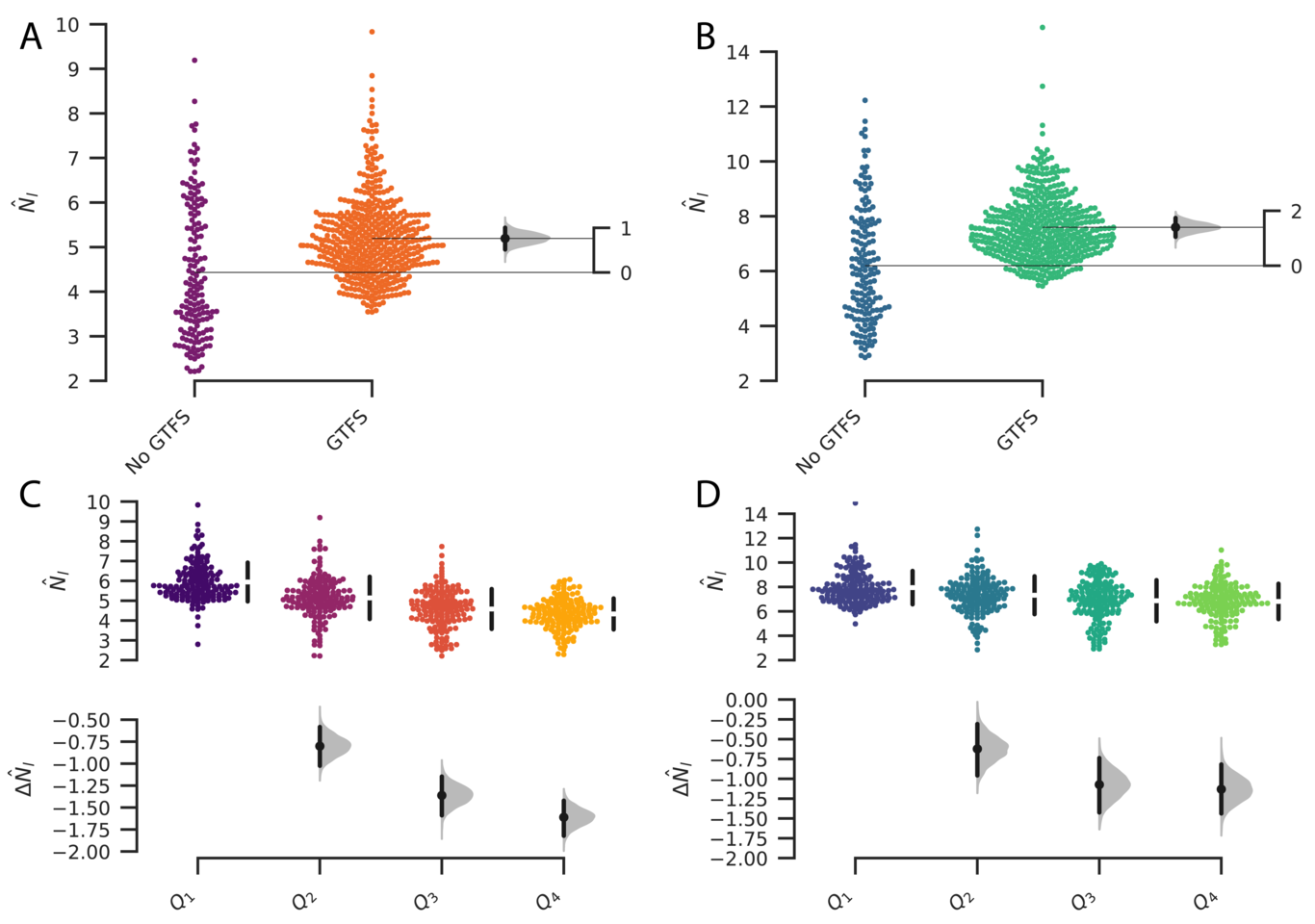}
\caption{{\textbf{How gender differences in mobility correlate with access to
public transport and socioeconomic status.}~Top: estimation plots of the
difference in the number of locations visited by women (\textbf{A}) and
men (\textbf{B}). where each~dot is a cell classified according to
having access to public transport (GTFS) or not (No
GTFS).\textbf{~}Bottom: estimation plots of the difference in the number
of locations visited by women (\textbf{C}) and men (\textbf{D}) where
each dot is a cell ranked by quartile of GSE ratio. Each group
represents cells in decreasing quartiles of income, from the top to the bottom quartile.
{\label{558897}}%
}}
\end{center}
\end{figure}\selectlanguage{english}

Access to different means of transportation plays a crucial role in determining individual mobility patterns. 
We investigate the relationship between the accessibility of public transport and the presence of gender disparities in mobility, by mapping the stops of the Santiago public transportation network (trams, buses and metro) onto the gendered mobility patterns of mobile phone users.  
To this aim, we characterize each cell of the SMR by the presence of at least one Transantiago stop, the public transport system that serves Santiago (see Materials and Methods). 
We then measure the gender effect on mobility through estimation statistics comparing two groups: the residents who have access to public transport and those who don't.

In Fig.~\ref{558897} each dot corresponds to the average number of locations, $\hat{N}_{l}$, visited by the residents of a given cell. 
Residents of cells with access to public transport (labeled as GTFS) visit significantly more locations than those living in areas without access to public transport (no GTFS). 
However, while access to public transport increases the mobility of both women (panel \textbf{A}) and men (panel \textbf{B}) in terms of number of unique locations visited, it does not fill the gender gap completely. 
The average value of $\hat{N}_{l}$ for women increases by $0.76$ ($95\%$ CI $[0.51,1]$) locations, when there is at least one Transantiago stop close to their home. 
For men, the mean value of $\hat{N}_{l}$ increases by $1.39$ ($95\%$ CI $[1.05,1.72]$) .   

One would expect the availability of public transport options to mainly serve the most deprived communities, whose residents must usually rely on public services to move around the city.  
We investigate the interplay between the accessibility of public transport and the socioeconomic fabric of different areas of Santiago by looking at the mean value of $\hat{N}_{l}$ in cells belonging to different quartiles of the GSE ratio distribution.  
Fig.~\ref{558897} shows the effect of the GSE ratio on $\hat{N}_{l}$ through estimation statistics, considering only those cells with access to public transport, grouped by quartiles of GSE ratio. 
Access to public transport proves to be less equalizing across socioeconomic segments for women (panel \textbf{C}) than for men (panel \textbf{D}). 
Women who live in cells belonging to the lowest quartile of the GSE ratio (Q4) visit on average $1.53$ ($95\%$ CI $[1.29,1.78]$) locations less than those in the highest quartile (Q1), even when they have access to public transport. 
The same difference measured for men is less than 1 location ($95\%$ CI $[0.51,1]$).

In some parts of the Santiago metro area residents may not have access to public transport, yet they may have access to a private vehicle, which might favor a higher mobility.
We assess the impact of owning a private vehicle on gender mobility by assigning each cell of Santiago to one of two groups, defined by census data: cells where there is more than 1 car every 4 residents, and cells where the number of private vehicles does not meet such threshold (see Materials and Methods). 
We then quantify the impact of belonging to the higher car ownership group through estimation statistics as done for the public transport analysis (see Fig.~S6 in the Supplementary Materials).
As expected, a higher proportion of car owners is associated to higher values of $\hat{N}_{l}$ both for women and men.
When looking at only those cells where having access to a car is more likely, we notice a smaller difference in the number of locations visited by the residents living in cells belonging to the different quartiles of GSE ratio for men than women.

\subsection{Gender differences in types of visited locations}
To further investigate gender differences in mobility beyond their most general statistical features, we turn our attention to potential gender differences in visitation patterns to different types of locations within the city. 
To characterize locations within the city, we turn to geographic databases of Points of Interest (POIs). We mostly use POI data from the openly accessible OpenStreetMap project (see Materials and Methods).
Although we have no information on whether the presence of a user near a specific POI actually corresponds to the user engaging with that POI, or whether the POI actually motivates the visit by the user, it is nevertheless of interest to study whether the visitation patterns to certain locations are gendered, and to this end the POI data we collected are potentially valuable as proxies for important characteristics of a given location (e.g., perceived or actual safety, pedestrian friendly areas, etc.) that might be associated with gender differences in visits. 

Thus, we study whether significant gender differences can be observed for visited locations that lie at or near a specific type of POI. 
To this aim, we compute, for each POI type, $k$, a POI density over the entire spatial domain under study, using a Kernel Density Estimator with varying bandwidth distance $d$. 
Then, we compute an average POI density ($\rho_k^{F, M}$) for all female and male users based on their set of visited locations. 
That is, $\rho_k^{F, M}$ is the density of POI type $k$ averaged over all locations (cells) visited by all females ($F$) or males ($M$).
Finally, we define the gender density ratio
\begin{equation*}
r_k = \rho^F_k \, / \, \rho^M_k \, ,
\label{eq:rho_ratio}
\end{equation*}
which, for each POI type $k$, is meant to indicate gender imbalances in visits to location associated with that specific POI type (see Materials and Methods for a complete description of the method).

\begin{figure}[th!]
\begin{center}
\includegraphics[width=\columnwidth]{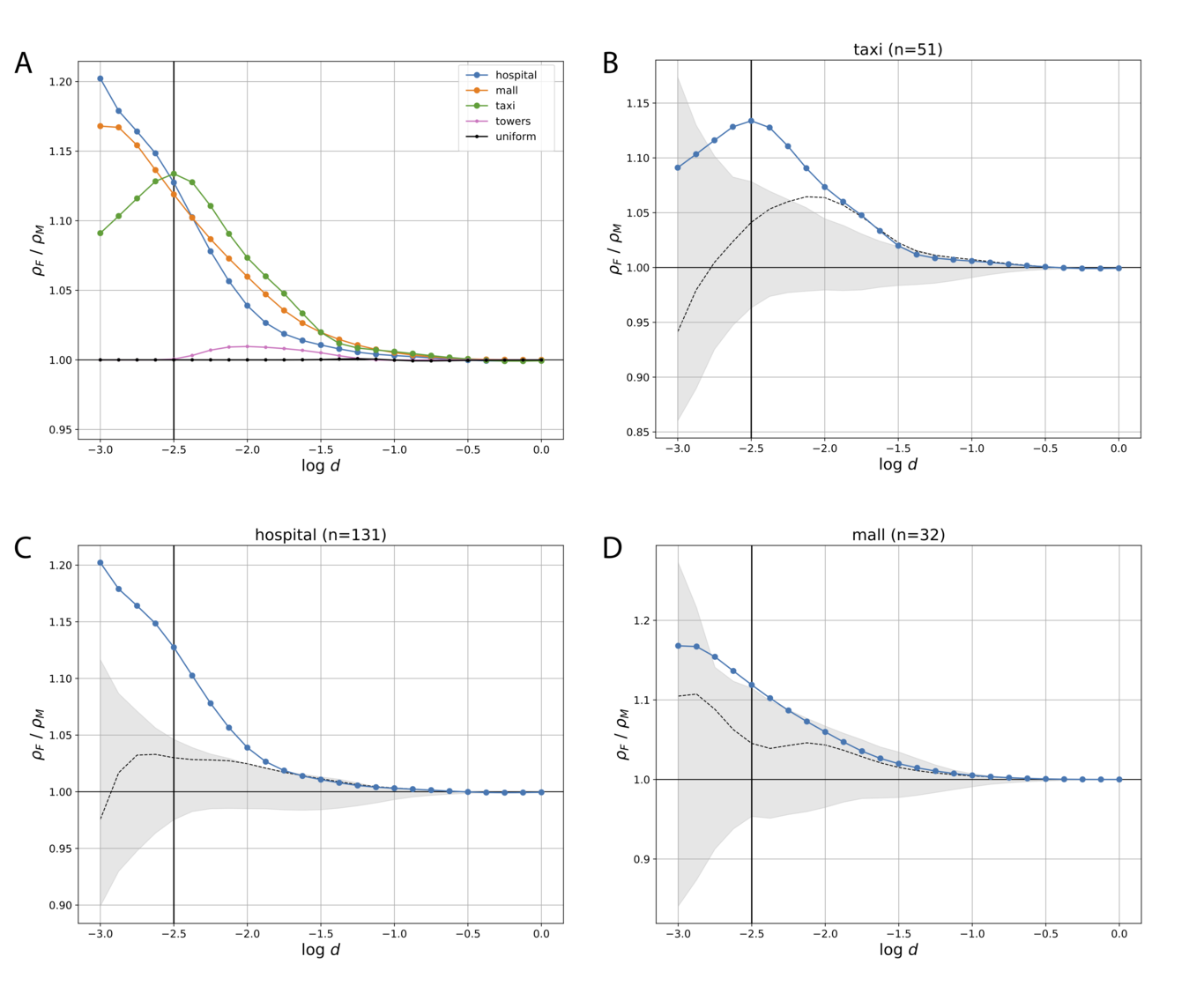}
\caption{{\textbf{Gender differences in visit patterns}.~Gender
ratio~\(\rho_{F}/\rho_{M}\) as~a function of the
bandwidth~\(d\)~ (decimal logarithm of the value in
kilometers), for a few selected POI types, including the "towers'` and
"uniform'' reference layers (\textbf{A}). The vertical line corresponds
to a kernel bandwidth comparable to the spatial resolution afforded by
the CDR data we use. The gender ratio~\(\rho_{F}/\rho_{M}\) as~a function
of the bandwidth~\(d\) is compared to two reference null
models for ``taxi'' (\textbf{B}), ``hospital'' (\textbf{C}) and ``mall''
(\textbf{D}). The shaded area indicates, for each value of the
bandwidth~\(d\), the 95\% confidence interval of the
values~\(\rho_{F}/\rho_{M}\) computed on the realizations of the spatial
sampling null model. The dashed line indicates the gender ratio computed
on the spatially perturbed null model.
{\label{213954}}
}}
\end{center}
\end{figure}

Fig.~\ref{213954}A shows the gender ratio $r_k$ as a function of the decimal logarithm of the bandwidth $d$ (in kilometers) for the three POI types with the largest imbalance in $r_k$: ``taxi'', ``hospital'' and ``mall''.
For comparison, Fig.~\ref{213954}A shows $r_k$ for two reference layers: the ``towers'' grid and the ``uniform'' grid, which are defined as uniform distributions of POIs (see Materials and Methods). 

We notice how for large distances the ratio $r_k \sim 1$, as the density for both females and males becomes essentially a Gaussian centered in the middle of the city. Moving towards shorter distances $d$, gender differences appear for some, but not for all, POI types indicating an imbalance of women visiting more frequently locations near some specific POIs (Fig.~S7 in the Supplementary Materials shows the values of $r_k$ for all POI types considered). 
The vertical line in Fig.~\ref{213954} corresponds to $\log_{10} d = -2.5$, for which more than 95\% of the kernel density is within $\pm 0.5$ km from the POI location (2 standard deviations on either side): hence this is the distance at which the spatial resolution of the kernel density estimator approaches the spatial resolution in the position of telephone towers, i.e., our spatial resolution limit on the position of users.
We also notice how the ``towers'' synthetic POI type exhibits some weak gender imbalance at the scale of a few kilometers, whereas the ``uniform'' reference layer yields $r=1$ for all values of $d$, as expected. 

To assess the significance of these gender imbalances, it is crucial to compare the observed values of $r_k$ against the values obtained by generating POI locations according to null models, or against values obtained by spatially perturbing the original POI locations.
Therefore, we compare the observed value with those obtained by using two different spatial null models that randomly generate synthetic POI locations (see Materials and Methods for details).
The first null model is based on randomly sampling phone towers according to the probability of visit by any user and then perturb locations so that they no longer lie on the regular grid.
The second null model is aimed at verifying the sensitivity of our results with respect to the specific set of locations of each POI type. 
It is built by generating a spatially-perturbed version of the same set of POIs, by randomly adding or subtracting an offset of $0.01$ from the latitude and longitude of each POI.

The results of this analysis are reported in Fig.~\ref{213954}, panels B, C and D, for the 3 POI types showing the largest gender imbalance.
The shaded area indicates, for each value of the bandwidth $d$, the $95\%$ confidence interval of the values $r_k$ computed on the realizations of the first null model with the same number of points $N_k$ as the original POI type.
For these POI types, the observed gender imbalances are strongly significant against the chosen null model for a broad interval of intermediate bandwidth values around $-2.5$.
The dashed lines describe the behavior of $r_k$ using the second null model, that is the spatially-perturbed version of the same set of POIs.
We notice how perturbing the POI locations pushes the ratio $r_{k}$ towards unity and into the $95\%$ confidence interval, making the deviation from unity insignificant for a broad range of bandwidth values $d$ (and in particular for $\log_{10} d \simeq -2.5$).
To further check the robustness of our results we compute the ratio $r_k$ considering different assumptions in the POI distributions and users' movements (see Figures S8, S9, S10 and Text E  in the Supplementary Materials). 
In all cases, the gender ratio $r_k$ appears to be strongly imbalanced for the same set of POI types, with females visiting more places near hospitals, malls and taxi stops.

\section{Discussion}

The main contribution of our study can be summarized as follows: women's travel patterns in the metropolitan area of Santiago are different than men's in various aspects, and such differences can be exposed by the analysis of large-scale anonymized mobile phone data disaggregated by sex. 

In their daily movements, women visit fewer locations than men and they are more localized, that is, they tend to distribute their time within a few preferred locations. 
Such reduced mobility for women might result from the interplay of cultural, infrastructure, resource, and safety constraints~\cite{kwan1999gender}. 
Although we mainly focused on capturing behavioral differences, we were able to relate mobility inequalities to a number of socio-demographic factors that may potentially explain the gender effect.
First, the gender inequality in mobility is widened in presence of income inequality. 
Indeed, a smaller gender gap in mobility characterizes the affluent municipalities of the Santiago Metropolitan Region.
More in general, the analysis of spatial patterns showed that income, employment and gender mobility equality are all positively correlated.
Our results confirm the value of mobile phone derived mobility metrics as a proxy for human development~\cite{pappalardo_analytical_2016, eagle2010network}. 
At the same time, we proved how the complex relations between gender, mobility and poverty can be elucidated by combining high-resolution telecommunication data with demographic statistics.

Transport and gender are tightly intertwined. 
By linking the gender gap in mobility with open data about public and private transport, we found that access to transport reduces mobility differences across socioeconomic segments for men but significantly less so for women.
Lower income remains a relevant factor that constraints women's mobility even when public transport is available, calling for more gender inclusive design policies for the transportation system of Santiago.
Finally, we found that not only women's mobility patterns in Santiago are different in terms of spatial and temporal features, but also in the \textit{type} of locations most frequently visited.
Indeed, we showed that it is possible to use Point-Of-Interest (POI) geographic databases to expose gender differences in the type of visited locations, as described by the presence or spatial proximity of specific points of interest. 
This allows to relate gender differences to categorized spatial features, suggesting hypotheses for further research as well as informing potential interventions. Specifically, we found that visits to cell towers close to \textit{hospitals}, \textit{malls} and \textit{taxi stands} are significantly gendered, and that this gender imbalance is strongly significant when contrasted to null models for POI spatial distribution and when tested against different assumptions on users mobility and POIs filtering (see Text E in the Supplementary Materials for a sensitivity analysis).
In summary, we found that visits to non-home, distant, non-frequently-visited (hence, likely, non-work) locations display significant gender differences for specific POI types, with females visiting more places near hospitals, malls and taxi stops.
This might indicate, in particular, that women carry a larger burden of caring for family members or related individuals in hospitals.
Our results thus demonstrate that mobile phone data are sensitive enough to capture the different mobility needs of men and women, related to their trips' purpose, thus representing a relevant source of information for urban planners to design gender responsive solutions. 

The use of mobile phone data to study gender mobility disparities - while promising - suffers from some limitations. 
One of them is the bias inherent to the data which can be related to two main issues: users' representativity and differences in calling activity. 
First, the users' sample might not be representative of the population under study: the sample size and composition will depend on the operator market share and in general users' demographics will not be fully representative of the population demographics. 
Second, mobility patterns are inferred through user's calling activity, which is known to be affected by age and gender, among other individual features~\cite{frias-martinez_gender-centric_2010}.
In our work we have controlled at our best for these biases, 
but we could not address some potential confounding effects, for instance those related to users' age, a variable not available to us.
To overcome the issue of a biased activity sampling for some users, the analysis of high-frequency x-Detail Records (XDR) could provide a valuable alternative.   
Also, when considering to extend our study to developing countries, we must mention biases in phone ownership, with women being more likely to use shared phones than men~\cite{blumenstock2010mobile}, although mobility estimates have been shown to be surprisingly robust to such biases at population level~\cite{wesolowski2013impact}.  

Our work falls within the growing research efforts to leverage large-scale data from digital traces to tackle crucial humanitarian and development questions. 
Previous studies have explored the use of big data such as transactions records, call detail records or social network penetration to map and investigate gender disparities~\cite{reed_observing_2016,lenormand_influence_2015,fatehkia2018using,garcia_analyzing_2018}.
While these novel data sources have proved their potential for social good, they also raise some important privacy concerns~\cite{jacques_mobile_2018, de_montjoye_unique_2013}.
We are aware of such concerns and we took them seriously into account by adopting a range of strategies to preserve users' privacy (see Materials and Methods).
Nevertheless, while recognizing the need of systematic and established approaches to the privacy-conscientious use of mobile phone data~\cite{de_montjoye_privacy-conscientious_2018}, we believe that their benefit can outweigh the risks, especially if analyzed in aggregrated form, as done here.    

Our study is the result of a joint collaboration between research centers, international organizations and private companies within the framework of a \emph{data collaborative}~\cite{susha2017data}.
We hope that our results will foster the creation of new data partnerships to further investigate the urban mobility experiences of women and girls within cities, so as to inform urban planners' decision-making process.

\par\null

\section{Materials and Methods}

{\label{sec:methods} }

\subsection{Mobile phone data and metadata}
{\label{sec:CDRdata} }

The mobile phone data set includes three months (May, June, July 2016) of anonymized (hashed) Call Detail Records (CDRs) enriched with gender, socioeconomic segment, and number of phone lines registered under that number for a total of 2,148,132,995 events (calls only). 
We filtered these events to those phone numbers that had only 1 registered line, placed at least one call per day on average (91 calls in total), had an identifiable home location, and visited more than two distinct locations over three months. This yielded a total of 418,624 unique users.
The associated socioeconomic segment was known for 315,844 users.
For all users, a binary gender value (F or M) was provided by the operator based on the information provided by the user at the time of subscription. In our sample, 51\% of the users are females. 

The socioeconomic status (GSE) of mobile phone users was classified by the phone carrier based on the GSE definition by the Chilean \selectlanguage{ngerman}\textit{Asociación de Investigadores de Mercado} (AIM) using a pay slip that cell phone users have to provide at the time of hiring the services of the telephone company. 
The socioeconomic groups are: Upper Class (AB), Wealthy Middle Class, (C1a), Emerging Middle Class (C1b), Typical Middle Class (C2), Medium-low Class (C3), Lower Class (D), Poors (E).
Since the GSE definitions can vary year by year, we divided our population into 2 socioeconomic segments corresponding to the upper class (AB,C1a,C1b,C2) and the lower class (C3, D, E). 
Our approach can be interpreted as dividing the population based on a reference household income, $I^*$, roughly corresponding to 1M Chilean pesos per year (1,500 USD).  
Then, for each spatial unit under study (tower cell or \textit{comuna}), we define the GSE ratio, that is the ratio between the resident population whose income is below the reference one, denoted as $P(I<I^*)$, and the population whose income is above $I*$, $P(I>I^*)$. 

We inferred users\selectlanguage{english}' mobility patterns by extracting the locations of each call placed or received, during the study period. Beside anonymization, to preserve the privacy of the users, the analysis of CDRs was carried out by spatially aggregating users\selectlanguage{english}' visited locations. 
More specifically, there are more than 1,300 towers (called Base Transceiver Stations, or BTSs) in urban Santiago for which we have access to the latitude and longitude rounded at the second decimal place. Due to this rounding, multiple antennas appear as having the same coordinates and they are merged into a single tower. 
This leads to a grid of 726 cells regularly spaced in which the center is given by the position (rounded) of the tower, and users\selectlanguage{english}' trajectories were aggregated at this spatial resolution.

We assigned a home location to each user based on user\selectlanguage{english}'s calling activity. 
We define a user\selectlanguage{english}'s home to be the most visited cell tower during the time interval 7pm - 8am, over the whole 91 days of data collection. Such approach, which can be denoted as ``time constrained home detection", has been shown to outperform several alternatives when tested on ground truth data~\cite{vanhoof2018performance}.

\subsection{Ethical considerations}
{\label{sec:ethics} }
We are aware of the concerns related to users' privacy when conducting research on CDRs data~\cite{de2018privacy}. 
This research was solely based on the analysis of anonymized data, after taking a number of precautions to ensure an appropriate protection of users' privacy and address any associated risks.
The mobile phone numbers of subscribers making and receiving calls was anonymized by the mobile phone operator inside their premises, through a hashing process using the secure SHA-3 algorithm.
Anonymized CDR data were never transferred outside of the operator\selectlanguage{english}'s system. 
Analysis of CDR data took place on the mobile operator\selectlanguage{english}'s systems and only the output of the analysis (aggregated population estimates and indicators) was subsequently made available to those researchers who were not based in Chile.
The analysis never singled out identifiable individuals and no attempts were made to link CDR data to third party data about an individual. No reports were made from towers that display fewer than three unique phone numbers and we did not work with antennas themselves, but aggregated all towers using rounded coordinates. Moreover, there is a natural privacy-preserving mechanism inherent in the data themselves: even at the finest level of granularity (the antenna), devices are never connected to the same antenna all the time, but rotate according to antenna demand (peak vs. regular times), time-of-day (some antennas get turned off at certain times), azimuth, etc. which makes extremely hard to identify a single user based on the location data that were analyzed in this work.

\subsection{Census data}
{\label{sec:census} }

The census data reports social, economic and demographic data about 17,574,003 people and 6,499,355 households in Chile collected by the \textit{Instituto Nacional de Estad\'isticas} (INE) in 2017 for the National Census (\textit{Censo de Poblaci\'on}). 
Data is publicly available at \url{https://www.censo2017.cl/}.
The census questionnaire is structured into three sections with questions about houses (\textit{Vivienda}), resident households  (\textit{Hogares}) and people belonging to each household (\textit{Personas}). 
Each questionnaire includes the location of the corresponding \textit{vivienda} at different administrative levels for the whole country. For our purposes we restricted our data analysis to the 52 \textit{comunas} of the Santiago Metropolitan Region.
For this study, in each \textit{comuna}, we extracted the following features that are reported in Table \ref{TableCorr}:
\begin{itemize}
\item the \textit{employment gender ratio} is defined as the ratio between (Employed Women/Total Women) and (Employed Men/Total Men);
\item the \textit{education gender ratio} is defined as the ratio between the proportion of women and men who completed a higher education level course: (High Edu Women/Total Women) and (High Edu Men/Total Men)
\item the \textit{general fertility rate} is defined as (Total Births / Total Women age 15-49) x 1,000;
\item \textit{couples households} is the number of households composed by the householder and a partner without children;
\item \textit{extended household} is the number of  households composed by a nuclear family plus other family members of the householder;
\item \textit{family household} is the number of  households composed by the householder, a partner and their children;
\item \textit{single parent household} is the number of  households composed by one householder and his/her children;
\item \textit{single person households} is the number of  households composed only by the householder.
\end{itemize}

\subsection{GTFS data}
{\label{sec:GTFS} }
The public transportation datasets of Santiago are  provided  in  General  Transit  Feed  Specification  (GTFS)  format. GTFS is a common format to represent public transportation  schedules. It is composed of a series of text files related to the routes and the stops of the  public transportation network. Here we consider only the file containing the GPS coordinates of the stops which enables us to identify cells of Santiago without access to public transportation stops.
The data is publicly available at \url{http://datos.gob.cl/dataset/33245/}.

\subsection{OpenStreetMap points of interest}
{\label{sec:poi_osm} }
Data about point of interests (POI) were downloaded from OpenStreetMap (OSM), a collaborative project to create a free editable map of the world, through the Overpass API\footnote{\texttt{https://wiki.openstreetmap.org/wiki/Downloading\_data}}. 
We collected OSM map features inside the CDR data boundaries, and filtered them considering only \textit{nodes} features, i.e. points with a geographic position, stored as pairs of latitude-longitude. We focused on those with the \texttt{amenity} tag, which are the most common type of POI, representing facilities used by visitors and residents.

In addition to the OpenStreetMap ``amenity'' POIs, we considered a few POI types from other data sources.
\begin{itemize}
\item \textbf{Malls} (\textit{``mall"} POI type).
This POI layer includes all tower locations $\vec{l}_i$ falling within the polygon describing the perimeter of mall complexes in the map of Greater Santiago (according to OpenStreetMap).

\item \textbf{Subway stops} (\textit{``metro''} POI type).
This POI layer includes all metro stop locations in the Metro Area of Santiago. Data is made publicly available by the \emph{Observatorio de Ciudades}, at the Faculty of Architecture of the Catholic University of Chile: \url{http://ideocuc.cl/maps/162/download}.

\item \textbf{Collective taxis} (\textit{``colectivos''} POI type).
This POI layer includes locations along the routes of shared taxis in Santiago (``colectivos''), obtained by scraping the geographical information used to draw taxi routes on the Web site~\url{http://www.ubicatucolectivo.cl/cliente\_final/all\_lines/vercion\_1.php?id=14}.
\end{itemize}

To aid with the interpretation of our results and to establish suitable baselines, we also create two synthetic POI types that we use as reference:
\begin{itemize}
\item \textbf{Tower clusters} (\textit{``towers''} synthetic POI type).
This POI type comprises all the locations of telephone towers (tower clusters) we use in our analysis. We know that towers are not uniformly distributed in space and that they are usually positioned following user density, hence their distribution contains information about the spatial distribution of the population in Santiago, and it is important to compare against this reference layer the results we obtain for other POI types, to rule out that the results are not purely determined by the spatial distribution of the population and of the towers.

\item \textbf{Uniform grid} (\textit{``uniform''} synthetic POI type).
This POI type contains all the vertices of the  two-dimensional spatial grid we use here (0.01 degree step long both the latitude and longitude coordinates), restricted to a rectangular area that covers the Greater Santiago Area. By design, this set of locations has no spatial structure (except for the size effects of the rectangular region for large distance).
\end{itemize}

\subsection{Gender differences in visit patterns}
We measured the excess ratio (women to men) of visits to locations characterized by the presence of specific category of POIs in the following way. 

For each POI type, we follow a standard non-parametric approach based on (multivariate) Kernel Density Estimation~\cite{Hwang1994} to compute a POI density over the entire spatial domain under study.
Namely, let us consider a given POI type $k$ comprising $N_k$ POIs at locations ${ \{ \vec{x}^{(k)}_1 ,  \vec{x}^{(k)}_2 , \ldots , \vec{x}^{(k)}_{N_k} \} }$, where the POI vector components are the latitudes and the longitudes of each POI. We define the density of POI $k$ at location $\vec{x}$ as
\begin{equation}
\rho_k(\vec{x}) = \frac{1}{N_k} \, \sum_{i=1}^{N_k}  K\left[  D( \vec{x} , \vec{x}^{(k)}_i  ) ; d  \right] \, , 
\label{eq:kde}
\end{equation}
where $K$ is a normalized non-negative kernel, the bandwidth $d > 0$ is a kernel parameter that defines the spatial scale over which the density distribution is smoothed, the function $D( \vec{x} , \vec{x}^{(k)}_i  )$ is the great circle distance of the positions vectors $\vec{x}$ and $\vec{x}^{(k)}_i $. 
We use a simple isotropic Gaussian kernel defined as
\begin{equation}
K(D; d) = \frac{1}{d \sqrt{2 \pi} } \, e^{-\frac{1}{2} (D / d)^2 }  \, ,
\label{eq:kde_kernel}
\end{equation}
and since we want to study potential gender differences at all spatial scales, we will not select a specific bandwidth value $d$, but rather carry out our analysis for a broad range of values, from slightly below the spatial resolution of our CDR data (hundreds of meters) to the size of the entire city (tens of kilometers).

Given a user $u$ and the POI densities defined above, we can define a user-level POI density by averaging over all locations visited by $u$:
\begin{equation}
\rho_k^{(u)} = \left< \rho_k(\vec{l}) \right>_{\vec{l} \in L(u)} \, .
\label{eq:poi_vector}
\end{equation}
That is, $\rho_k^{(u)} $ is the density of POI type $k$ averaged over all locations $L$ visited by user $u$.
Notice that this density has an implicit dependency on the bandwidth parameter $d$ of the kernel density estimator.

Finally, we carry out averages of the user-level POI densities separately over all males and female.
That is, indicating with $U_F$ and $U_M$ the set of all female and male users, respectively, we define:
\begin{equation}
\rho^F_k = \left< \rho_k^{(u)}  \right>_{u \in U_F} \, ,
\label{eq:rho_f}
\end{equation}
and similarly
\begin{equation}
\rho^M_k = \left< \rho_k^{(u)}  \right>_{u \in U_M} \, .
\label{eq:rho_m}
\end{equation}
Finally, we define the gender density ratio
\begin{equation}
r_k = \rho^F_k \, / \, \rho^M_k \, ,
\end{equation}
which, for each POI type $k$, is meant to indicate gender imbalances in visits to location associated with that specific POI type.
This ratio also has an implicit dependency on the bandwidth parameter $d$ used for kernel density estimation,
hence we study $r_k$ as a function of both the POI type $k$ and of the smoothing distance $d$.

To assess statistical significance, we compare the observed value of $r_k$ with those obtained by using two spatial null models that randomly generates synthetic POI locations with two methods:
\begin{itemize}

\item \textbf{Spatial sampling} Given a POI type $k$ that comprises $N_k$ POIs, we generate a realization of the null model by sampling (with repetition) $N_k$ towers according to the probability $p_i$ that a tower $i$ is visited by any user, i.e., by the fraction of CDR records that tower (tower cluster) is associated with. Subsequently, we perturb the locations so that they no longer lie on the regular grid, by adding a uniformly distributed random variate in the range $[-0.01, +0.01]$ to both the latitude and the longitude. For each POI type $k$ we generate 100 realizations and compute the ratio $r_k$ for each realization. Finally, we compute the $95\%$ confidence interval for the distribution of $r_k$ generated by the null model.

\item \textbf{Spatial perturbation} For each POI type $k$ we build a spatially-perturbed version of the same set of POIs by randomly adding or subtracting an offset of $0.01$ from the latitude and longitude of each POI. We treat the perturbed POIs as a new POI type $k^\prime$ and compute the gender ratio $r_{k^\prime}$ for all values of $d$, as described above.
\end{itemize}

\subsection{Mobility metrics}
{\label{sec:metrics} }
We analyzed user's mobility computing four different mobility metrics, first defined at individual level and then aggregated by averaging over users with same gender, GSE or home location. 

As a basic measure of mobility behavior, we computed the number of distinct locations visited by a user, $N_l$, that corresponds to the number of distinct grid cells in which a user made or received at least one call over the 3 month period. 
Given that $N_l$ can vary significantly between users and it is affected by fluctuations of user's activity, we also computed the ``core activity locations", $\hat{N}_l$, defined as the set of locations that account for 80\% of a user's calling activity. 

To quantify the diversity of individual mobility, we calculated the Shannon entropy of user's trajectories as:
\begin{equation}
S=-\sum\limits_{l\in L}p_l\ln p_l \,,
\end{equation}
where $L$ is the full set of locations visited by a user, and $p_l$ is the the probability of observing a user in $l$, computed as the fraction of calls made by the user at location $l$. 
A user with high $S$ will distribute her visits across many different locations with equal probability, while a lower $S$ corresponds to a higher regularity of mobility patterns with a smaller set of regularly visited locations~\cite{Song_2010}.

Finally, we measured the radius of gyration of each user, $r_g$, which quantifies the characteristic distance traveled by an individual. 
It is defined as:
\begin{equation}
r_g=\frac{1}{L} \sqrt{\sum\limits_{i=1}^{L}(\mathbf{r_i}-\mathbf{r_{cm}})^2}
\end{equation}
where $L$ is the full set of locations visited by a user, $\mathbf{r_i}$ is the vector of coordinates of location $i$ and $\mathbf{r_{cm}}$ is the vector of coordinates of the center of mass, weighted by the visiting frequency $p_i$.

For each mobility metric, we carry out averages of the user-level metrics separately over all males and females who live in a given location $l$ (\emph{comuna} or cell).
That is, indicating with $U_{F,l}$ and $U_{M,l}$ the set of all female and male users who live in $l$, respectively, we define the average of the mobility metric $x$:
\begin{equation}
x_F = \left< x^{(u)}  \right>_{u \in U_{F,l}} \, ,
\end{equation}
and
\begin{equation}
x_M = \left< x^{(u)}  \right>_{u \in U_{M,l}} \, .
\end{equation}
Finally, to measure the gender gap for a given mobility metric $x$, we measure the gender ratio:
\begin{equation}
R_{x} = x_F / x_M \, .
\end{equation}

\textbf{Acknowledgments} : We thank Telefonica I+D for providing access to the data and the infrastructure to analyze it, in particular Pablo Garcia Briosso. 
We thank Eduardo Graells-Garrido for help with data visualization.

\textbf{Funding:} This project was funded in part by the Data2X Initiative through the Big Data for
Gender Challenge Awards. The authors acknowledge financial support from Movistar - Telefonica Chile, the Chilean government initiative CORFO 13CEE2-21592
(2013-21592-1-INNOVA\_ PRODUCCION2013-21592-1), Conicyt PAI Networks
(REDES170151) ``Geo - Temporal factors in disease spreading and
prevention in Chile''. 
The support of the CRT Foundation through the Lagrange Project is gratefully acknowledged.
The funders had no role in study design, data collection and analysis, decision to publish, or preparation of the manuscript.

\textbf{Competing interests:} The authors declare no competing
interests.

\newpage

{\label{660096}}\selectlanguage{english}
\begin{table*}
\caption{{\label{Tablediff} Estimation statistics of gender differences in mobility, disaggregated by users' socioeconomic group (GSE). GSE are ranked from the highest (ABC1) to the lowest (E). For each metric $X$, the effect size $\Delta$ is measured as $\Delta X = \langle X \rangle_{M} - \langle X \rangle_{F}$. Square brackets report 95\% CIs.}}
\begin{center}
\begin{tabular}{rcccc}
     &  $\Delta S $ & $\Delta \hat{N}_l$ & $\Delta N_l$ & $\Delta r_G$\\ 
    \hline
ABC1& 0.19 [0.17 - 0.20]  & 1.66 [1.53 - 1.79] & 7.28 [6.72 - 7.82] & 0.83 [0.75 - 0.90] \\
C2 & 0.23 [0.22 - 0.24] &  2.08 [1.99 - 2.18] & 8.54 [8.14 - 8.95] & 0.94 [0.88 - 1]\\
C3& 0.28 [0.27 - 0.29] &  2.50 [2.40 - 2.59] &  10.16 [9.78 - 10.53] & 1.15 [1.09 - 1.21]\\
D& 0.30 [0.30 - 0.31] &  2.53 [2.45 - 2.61]& 10.22 [9.87 - 10.53] & 1.26 [1.21 - 1.32]\\
E& 0.31 [0.29 - 0.33] & 2.50 [2.32 - 2.66] & 9.82 [9.16 - 10.48] & 1.36 [1.25 - 1.47]\\
\end{tabular}
\end{center}
\end{table*}\selectlanguage{english}

\begin{table*}
\caption{{\label{TableCorr} Semi-partial correlation values (Pearson) between $R_{S}$ and $R_{\hat{N}_l}$ and the sociodemographic features of 51 municipalities in the SMR. All correlation values are computed by correcting for differences in calling activity and users distribution by gender. We report statistical significance using the following notation: * = $p<0.05$, ** = $p<0.01$, *** = $p<0.001$.}}
\begin{center}
\begin{tabular}{rcc}
     &  $R_{S}$ & $R_{\hat{N}_l}$\\ 
    \hline
{\bf GSE ratio (log)} & -0.59***  & -0.53***  \\
{\bf    HDI} & 0.42** &  0.37** \\
{\bf    education gender ratio}  & -0.08 &  -0.10 \\
{\bf    employment gender ratio}  & 0.51*** &  0.37**  \\
{\bf    general fertility rate}  & -0.53*** &  -0.40**  \\
{\bf    couples household} & 0.55*** & 0.50*** \\
{\bf    extended household}  & -0.61***  &  -0.57*** \\
{\bf    family household}  & -0.30 & -0.14 \\
{\bf    single parent household}  & -0.32*  & -0.32* \\
{\bf    single person household}  & 0.56***  &  0.44** \\
\end{tabular}
\end{center}
\end{table*}

\newpage

\selectlanguage{english}
\bibliographystyle{Science}

\newpage

\begin{center}
 \textbf{\large Gender gaps in urban mobility\\Supplementary Material}\\[.2cm]
Laetitia Gauvin,$^{1}$ Michele Tizzoni,$^{1}$, Simone Piaggesi$^{1,2}$, Andrew Young$^3$, Natalia Adler$^4$, Stefaan Verhulst$^3$, Leo Ferres$^{5,6}$ and Ciro Cattuto$^1$\\[.1cm]
 {\itshape ${}^1$ISI Foundation, Torino, Italy\\
 ${}^2$Doctoral School in Data Science and Computation, University of Bologna, Bologna, Italy\\
 ${}^3$The Governance Lab, New York University, New York, NY, United States\\
${}^4$United Nations International Children's Emergency Fund (UNICEF), New York, NY, United States\\
${}^5$Data Science Institute, Universidad del Desarrollo, Santiago, Chile\\
${}^6$Telefónica R\&D, Santiago, Chile\\
}
\end{center}

\setcounter{equation}{0}
\setcounter{figure}{0}
\setcounter{table}{0}
\setcounter{page}{1}
\renewcommand{\theequation}{S\arabic{equation}}
\renewcommand{\thefigure}{S\arabic{figure}}
%
%
%
%
%
\newpage

\begin{figure}[h!]
\begin{center}
 
 \includegraphics[width=0.9\linewidth]{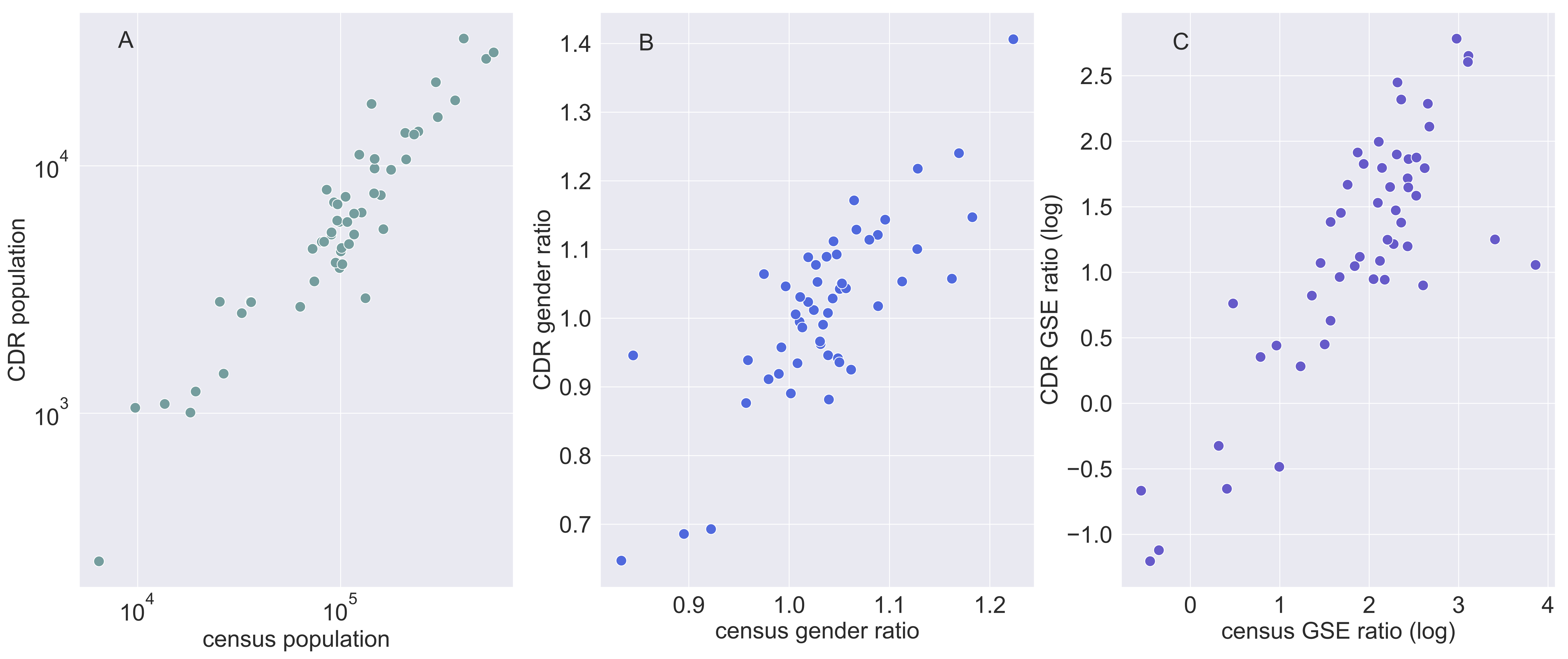}
 
 \caption{a) Scatter plot of the gender ratio measured on the CDR data by municipality with respect to the census gender ratio. b) Scatter plot of the ratio of the poorest population (according to the GSE ratio) over the weathiest population by municipality measured computed from the survey with respect to the one measured using the metadata associated to the CDR data.
  The CDR population sample is highly representative of the population distribution as confirmed by the high correlations of $0.93$, $0.79$, $0.88$ measured between the CDR data and the baseline (census or survey) respectively for the population, the gender ratio and the GSE ratio. We remind that the  GSE population ratio corresponds to the ratio of the population having an income below  a reference income $P(I<I^*)$ over the population having an income below this reference income $P(I>I^*)$. The repartition of the socio-economic groups in each \textit{comunas} has been extracted using the survey \textit{Encuesta Origen Destino de Viajes} (EOD) realized by the \textit{Secretaría de Planificación de Transporte} (SECTRA) between July 2012 and November 2013, by assignment of Chilean Ministry of Transport and Telecommunications. The dataset contains the income of $60,054$ individuals ($47\%$ males, $53\%$ females). 
Socio-economic groups were assigned to each household of the survey based on their income using the definition of the socio-economic groups as given by \url{http://www.emol.com/noticias/Economia/2016/04/02/796036/Como-se-clasifican-los-grupos-socioeconomicos-en-Chile.html}
}
 \end{center}
 {\label{Rep}}
 \end{figure} 


\begin{figure}[h!]
\begin{center}
 \includegraphics[width=0.5\linewidth]{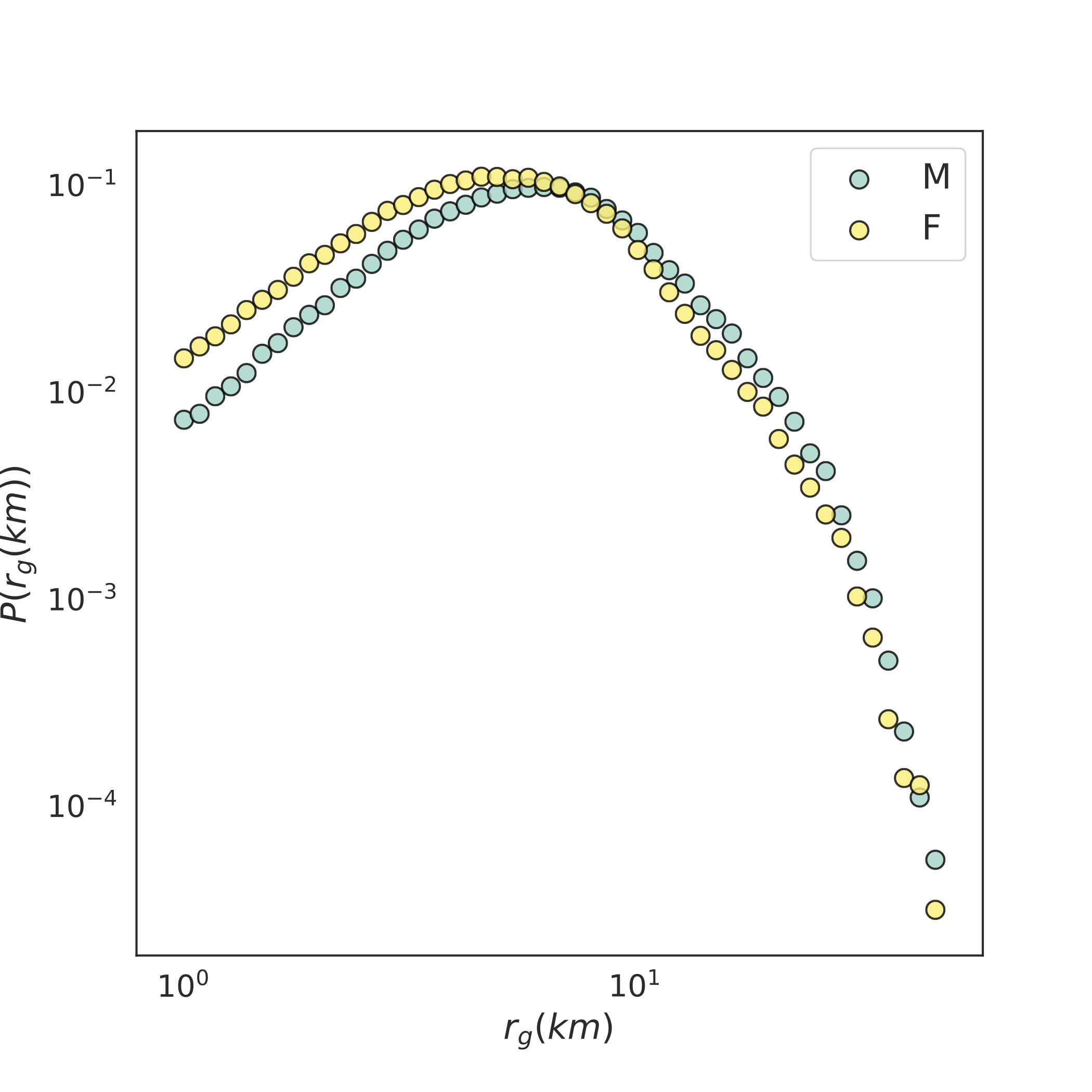}
 \caption{Distribution of the radius of gyration ($r_g$) disaggregated by gender. 
 The $r_g$ distribution spans a similar range of distances for women and men, however women tend to have a smaller radius of gyration.}

 \end{center}
 {\label{Rog}}
 \end{figure}

\newpage

\begin{figure}[h!]
\begin{center}
 \includegraphics[width=0.6\textwidth]{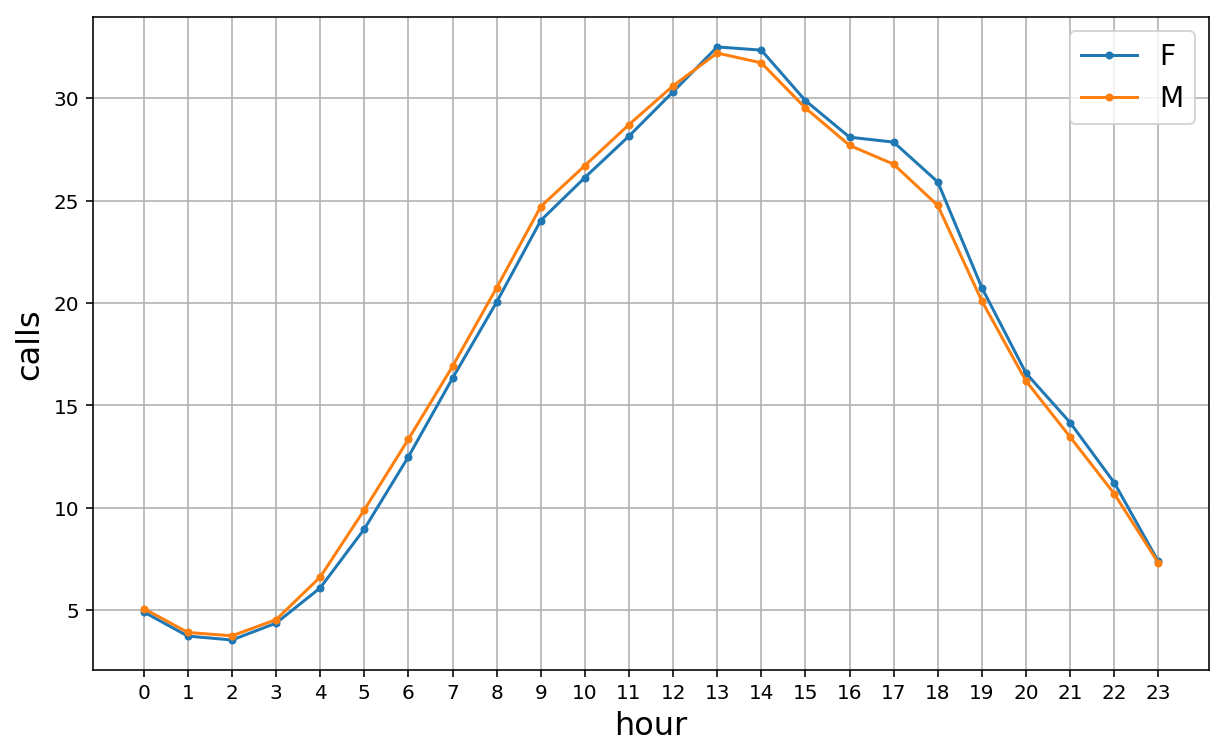}
 \caption{Distribution of the average number of calls made or received by the users during the day, disaggregated by gender. 
 The average is computed hourly over the three month period of our study, and then normalized by the total average calling activity for males and females respectively.}
 \end{center}
 {\label{daily}}
 \end{figure}

\newpage
%



\newpage

\begin{figure}[h!]
\begin{center}
 \includegraphics[width=0.6\linewidth]{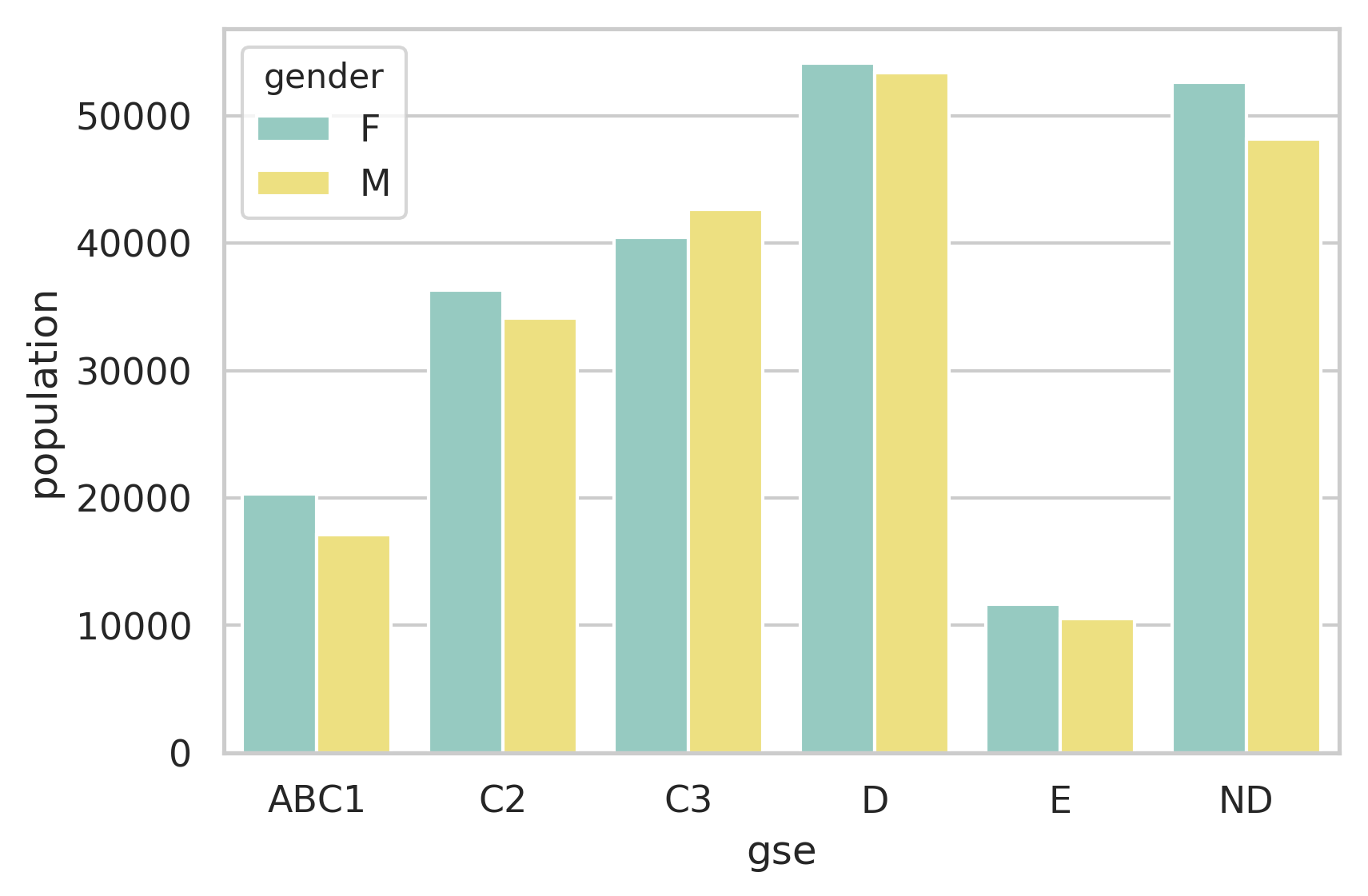}
  \caption{Distribution of the CDR users by GSE and by gender. There are 5 GSE classes from the highest to the lowest level: ABC1, C2, C3, D, and E. Users in the class ND have no GSE assigned. Overall, we observe that both genders are generally equally represented across socio-economic segments.}

 \end{center}
 {\label{GSE-distri}}
 \end{figure}

\begin{figure}[h!]
\begin{center}
{\label{fig:gender_mob_eco_ratio}}%
\includegraphics[width=0.9\linewidth]{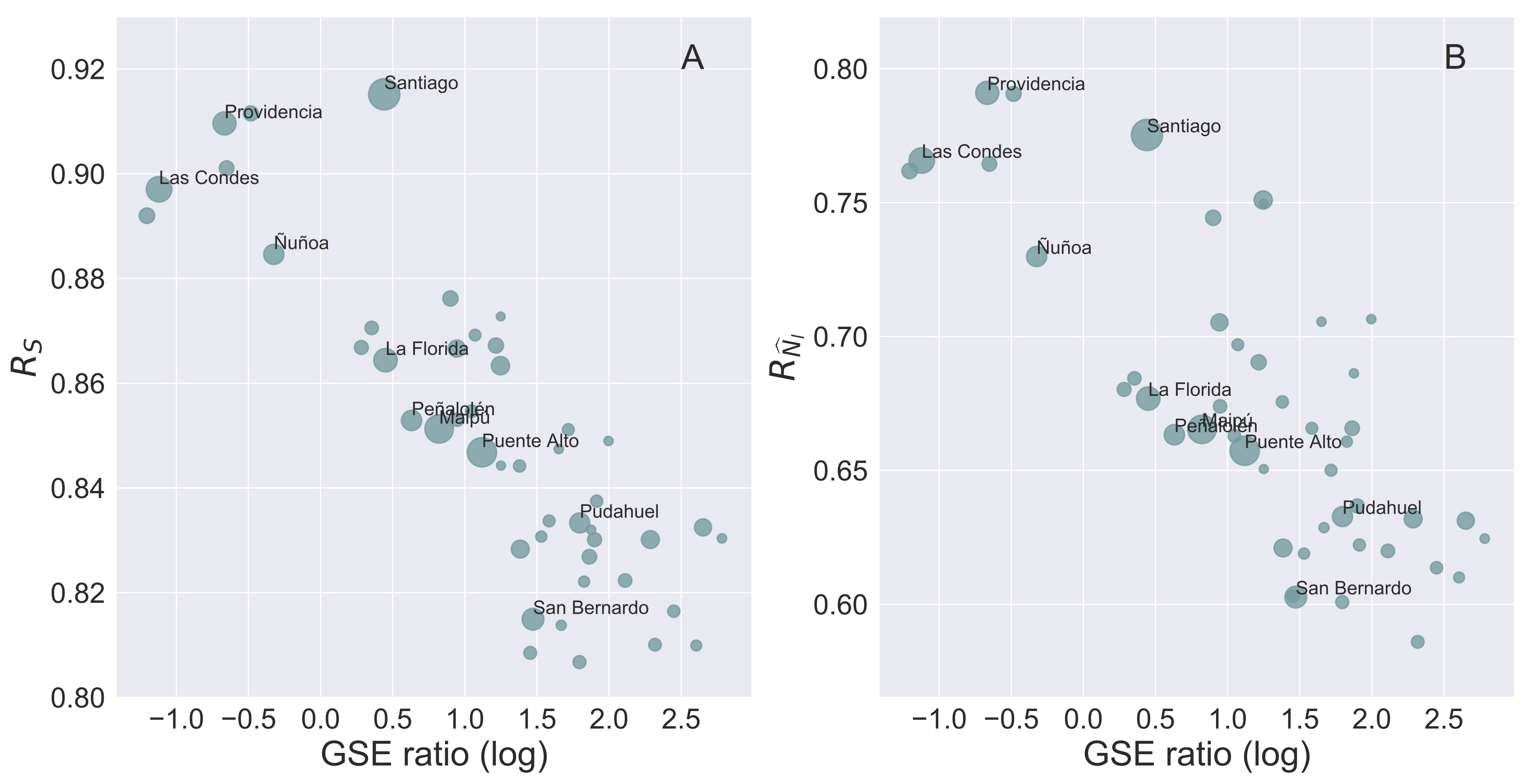}
\caption{{\textbf{Scatterplots of the mobility gap against the GSE ratio by Comuna
of the Santiago Metropolitan area}. The mobility gap is measured by the
entropy ratio (\textbf{A}) and the women to men ratio of the number of
visited locations that correspond to at least $80\%$ of the activity of each user (\textbf{B}).The size of the points is proportional to the population of the \textit{comunas}.
}}

\end{center}
\end{figure}

\newpage

 
\begin{figure}[h!]
\begin{center}
\includegraphics[width=1.00\columnwidth]{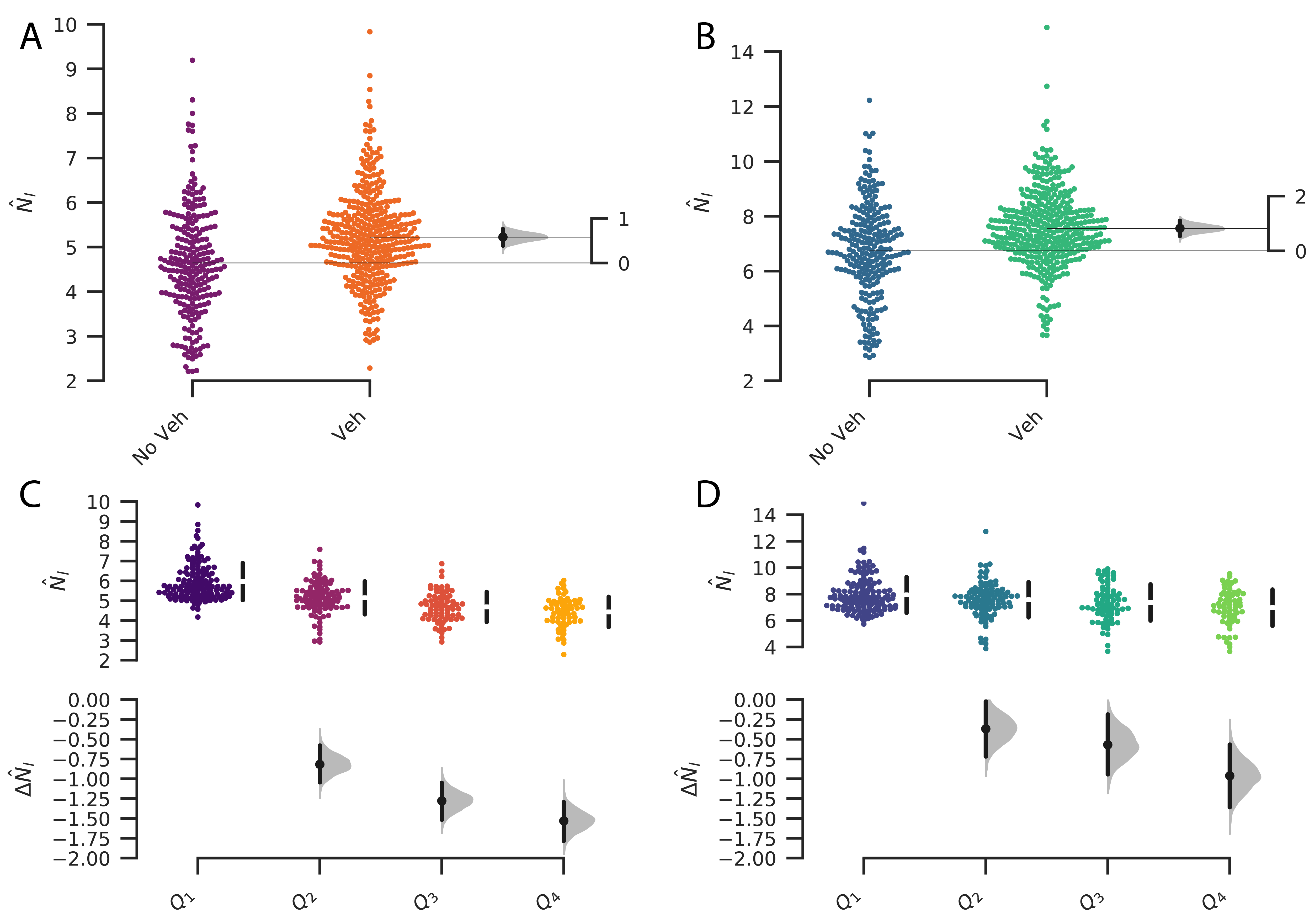}
\caption{{\textbf{Estimation plots of the average difference in number of
locations visited by women and men by cells classified according to
access to private transportation and socio-economic groups.} Estimation
plots of the difference in number of locations visited by women
(\textbf{A}) and men (\textbf{B}) under the conditions of living in
cells where the average number of cars is below~ (No Veh) or above~
(Veh) 1 over 4 per person\textbf{.~}Estimation plots of the difference
in number of locations visited by women (\textbf{C}) and men
(\textbf{D}) cells where the average number of cars isabove1 per 4
person by quartile of GSE ratio. Each group represents cells in
decreasing quartiles of income, from the top 25\% to the bottom 25\%. 
As for public transportation, we observe an increase in mobility  both for women and mean when their probability of having access to a private vehicle is higher.If we focus on cells where having access to car is more likely (i.e. where the average number of cars is above the threshold we introduced), we notice a smaller difference in the number of locations visited by the residents living in cells belonging to the different quartiles of GSE ratio for men than women.
Overall the access to transportation either public or private is linked to higher mobility but it does not fill the gender gap observed for the poorest areas of Santiago.
{\label{767582}}%
}}
\end{center}
\end{figure}

\clearpage

\begin{figure}[h!]
\begin{center}
\includegraphics[width=1.00\columnwidth]{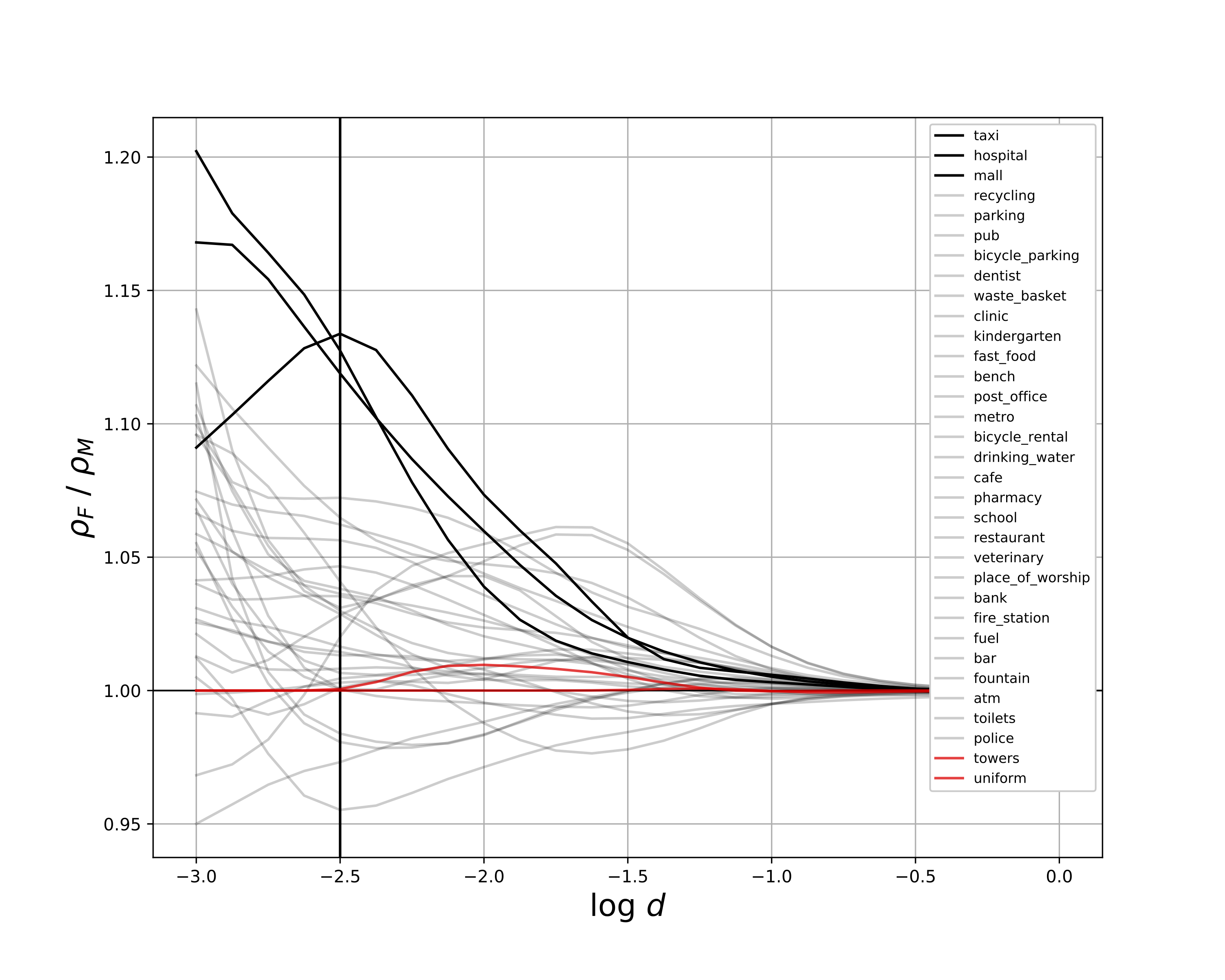}
\caption{{\textbf{Gender differences in visit patterns.} Gender
ratio~\(\rho_{F}/\rho_{M}\) as~a function of the
bandwidth~\(d\)~ (decimal logarithm of the value in
kilometers), for all POI types considered, including the ``towers'' and
``uniform'' reference layers.  
The vertical line corresponds
to a kernel bandwidth comparable to the spatial resolution afforded by
the CDR data we use.
The ratio for the three POI types showing the largest gender imbalance at $log(d) = -2.5$ (``taxi'', ``hospital'', and ``mall'') is highlighted in black.} %
}
\end{center}
\end{figure}

\begin{figure}[h!]
\begin{center}
\includegraphics[width=1.00\columnwidth]{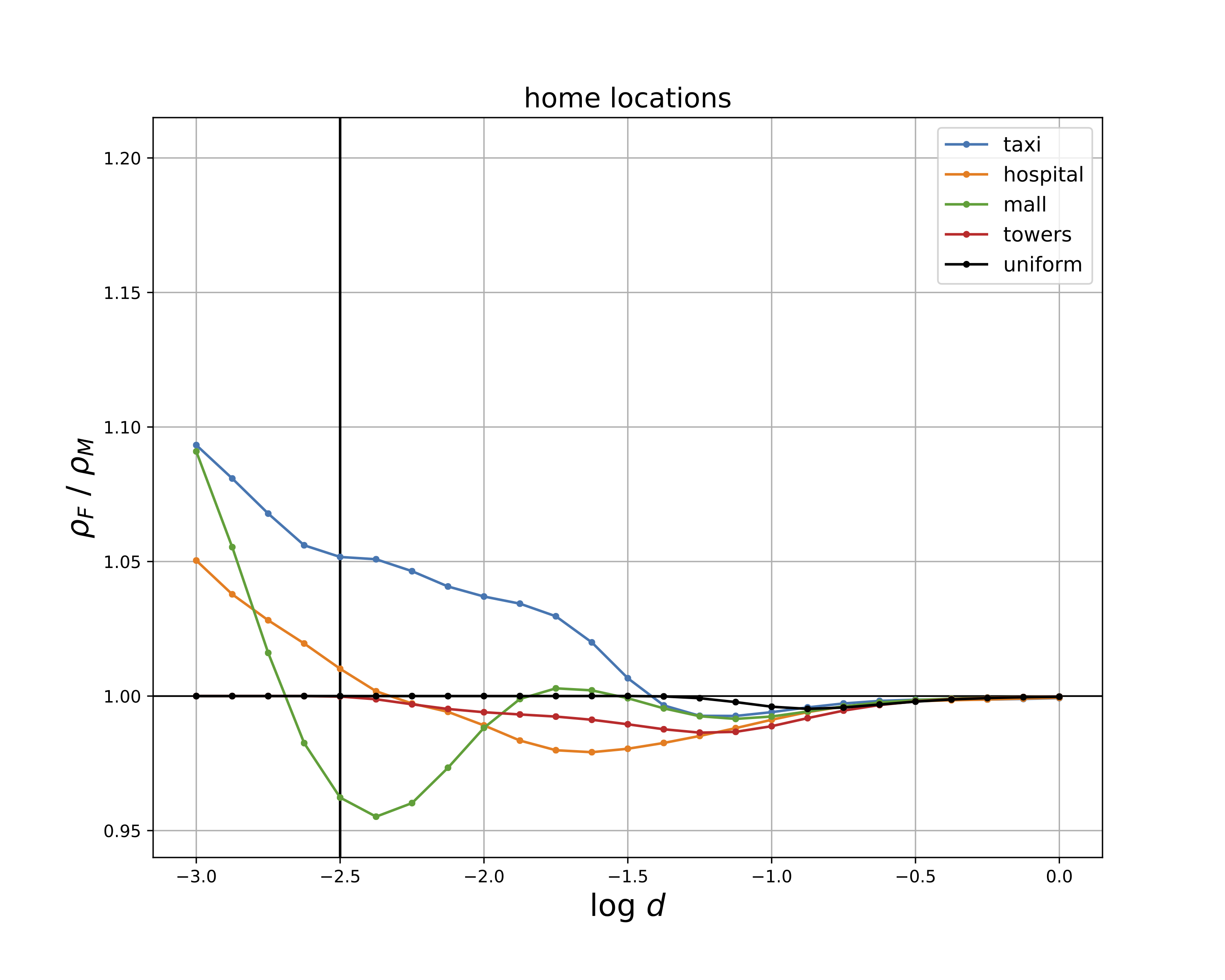}
\caption{{\textbf{Gender differences in visit patterns: sensitivity to home location.} 
Gender ratio~\(\rho_{F}/\rho_{M}\) as~a function of the
bandwidth~\(d\)~ (decimal logarithm of the value in
kilometers), for the three POI types showing the largest gender imbalance (``taxi'', ``hospital'',  and ``mall'') 
and the ``towers'' and ``uniform'' reference layers.  
The vertical line corresponds
to a kernel bandwidth comparable to the spatial resolution afforded by
the CDR data we use. 
The gender ratio is computed by considering only home locations as the set of visited locations of each user.
The gender ratio drops significantly close to or below 1 for all the three POIs (``taxi'', ``hospital'',  and ``mall''). This means that the gender imbalance observed in the main analysis for all non-home locations is larger than we could expect if only based on the differences in home locations by gender.}}
\label{SI_POI_home}
\end{center}
\end{figure}

\begin{figure}[h!]
\begin{center}
\includegraphics[width=1.00\columnwidth]{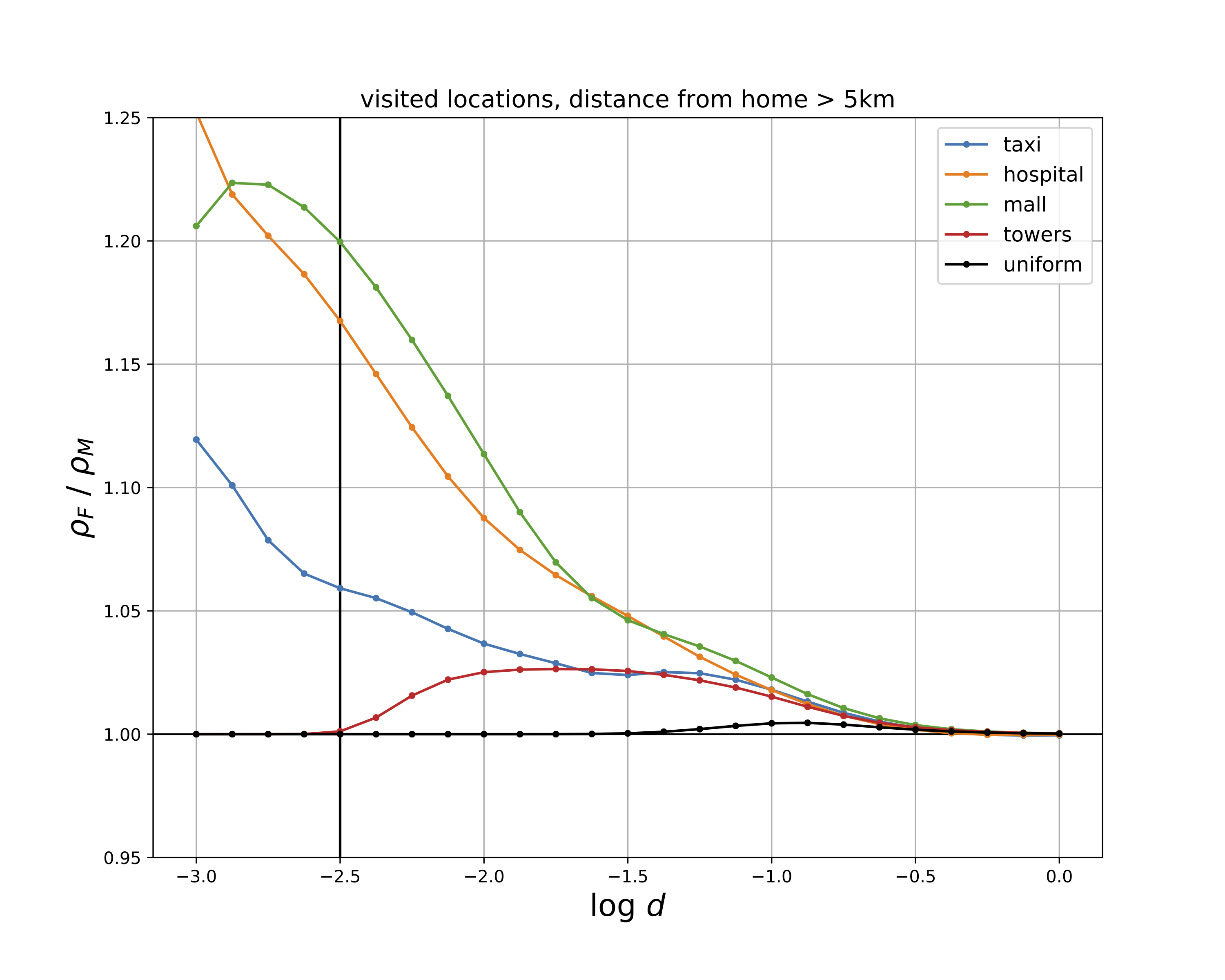}
\caption{{\textbf{Gender differences in visit patterns: sensitivity to distance from home.} 
Gender ratio~\(\rho_{F}/\rho_{M}\) as~a function of the
bandwidth~\(d\)~ (decimal logarithm of the value in
kilometers), for the three POI types showing the largest gender imbalance (``taxi'', ``hospital'',  and ``mall'') 
and the ``towers'' and ``uniform'' reference layers.  
The vertical line corresponds
to a kernel bandwidth comparable to the spatial resolution afforded by
the CDR data we use. 
The gender ratio is computed by excluding from the set of visited locations of each user all locations that fall within a 5 km distance from home.} %
}
\label{SI_POI_distant}
\end{center}
\end{figure}

\begin{figure}[h!]
\begin{center}
\includegraphics[width=1.00\columnwidth]{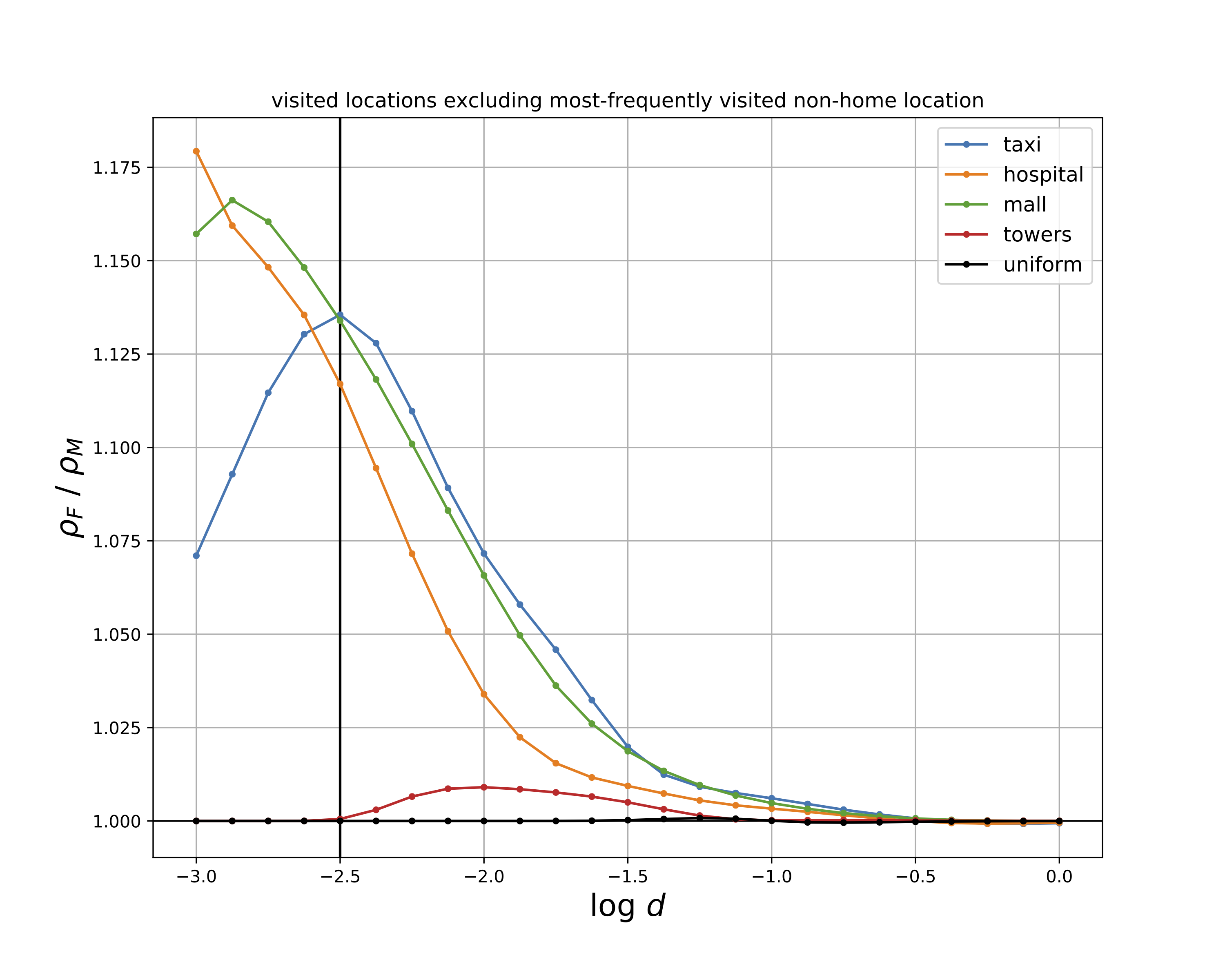}
\caption{{\textbf{Gender differences in visit patterns: sensitivity to work location.} 
Gender ratio~\(\rho_{F}/\rho_{M}\) as~a function of the
bandwidth~\(d\)~ (decimal logarithm of the value in
kilometers), for the three POI types showing the largest gender imbalance (``taxi'', ``hospital'',  and ``mall'') 
and the ``towers'' and ``uniform'' reference layers.  
The vertical line corresponds
to a kernel bandwidth comparable to the spatial resolution afforded by
the CDR data we use. 
The gender ratio is computed by excluding from the set of visited locations of each user both the first the second most frequently visited locations (home and work).} %
}
\label{POI_work}
\end{center}
\end{figure}

\begin{figure}[h!]
\begin{center}
\includegraphics[width=1.00\columnwidth]{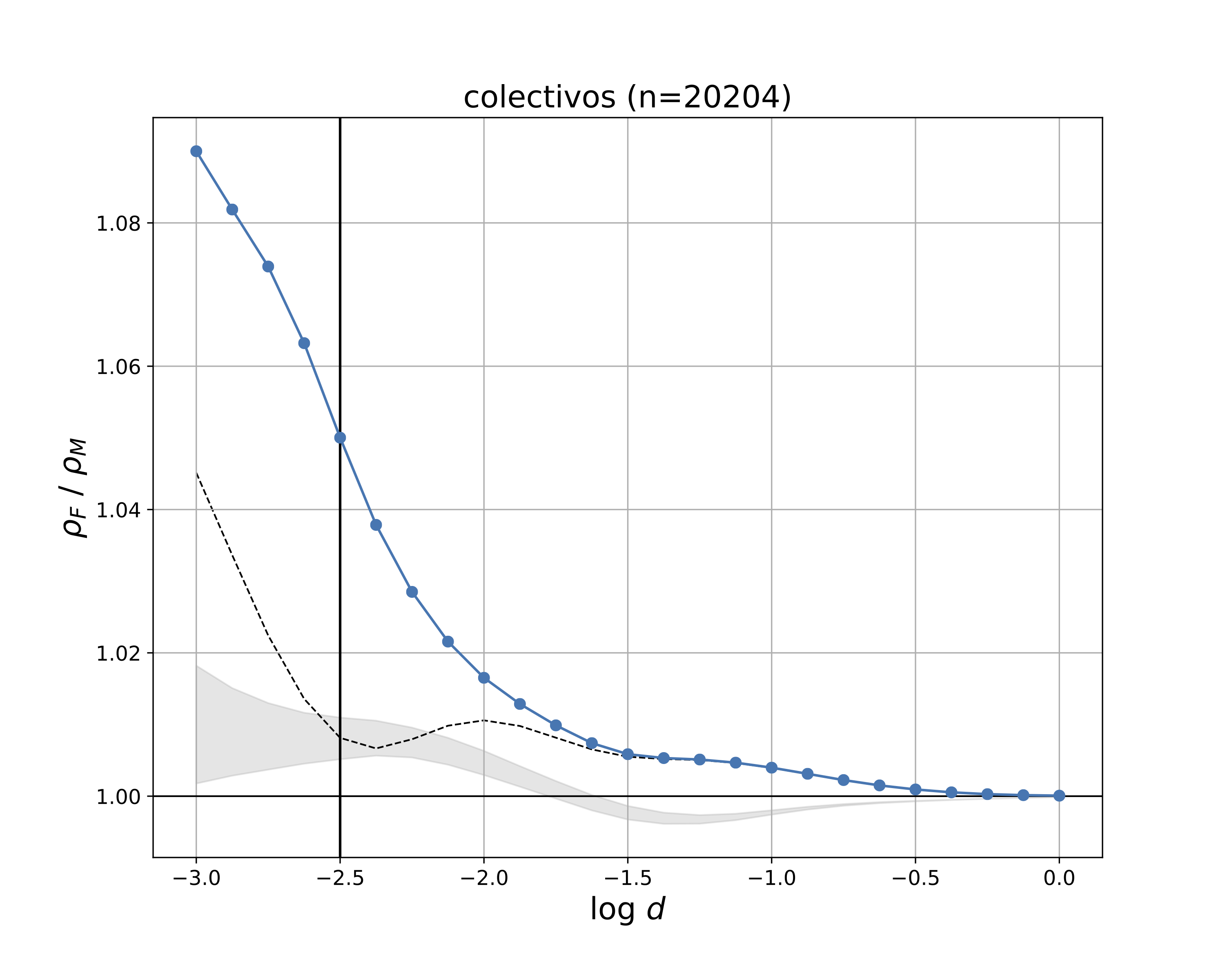}
\caption{{\textbf{Gender differences in visit patterns: collective taxis.} 
Gender ratio~\(\rho_{F}/\rho_{M}\) as~a function of the
bandwidth~\(d\)~ (decimal logarithm of the value in
kilometers), for the ``colectivos'' (collective taxis) POI type.
This POI layer includes locations along the routes of shared taxis in Santiago (``colectivos''), obtained by scraping the geographical information used to draw taxi routes on the Web site~(\url{www.ubicatucolectivo.cl/cliente\_final/all\_lines/vercion\_1.php?id=14}).
The vertical line corresponds
to a kernel bandwidth comparable to the spatial resolution afforded by
the CDR data we use. 
The shaded area indicates, for each value of the bandwidth~\(d\), the 95\% confidence interval of the values~\(\rho_{F}/\rho_{M}\) computed on the realizations of the spatial sampling null model. The dashed line indicates the gender ratio computed on the spatially perturbed null model.
}
}
\label{POI_colectivos}
\end{center}
\end{figure}

\clearpage

\section*{Text A: Controlling for differences in call activity by gender}
We investigated the relationship between socio-demographic features and the gender differences in mobility by computing the Pearson semi-partial correlation coefficient between any two types of variables, controlling both for call activity and population size. 
For completeness, we also performed a sensitivity analysis of our results with respect to two types of sampling of the users' call activity distribution. 
First, we down sampled the call activity of men by randomly removing $25\%$ of calls (resp. $50\%$) for each male user. 
Then, we computed the average Nemenman-Shafee-Bialek (NSB) entropy estimator \cite{nemenman2002entropy} for women, $NSB_F=1.7$ and men $NSB_M=1.87$ (resp 
$NSB_F=1.7$ and  $NSB_M=1.94$). 
We see that even after a drastic down-sampling of men's call activity, the entropy of women is still smaller on average. 
Second, we selected only users who made at least $200$ calls during the three month time period of our study. This corresponds to restricting our sample to $72\%$ of the total number of users. 
Then, we selected $200$ calls at random for each user and recompute the discrete entropy estimator so to compare users with the same total activity. 
In this case, the measured entropy estimators are $NSB_F=1.73$ and  $NSB_M=2.01$, with women still displaying a smaller average entropy.
In both cases, we observe a difference between the male and female sample that is statistically significant according to a Kruskal-Wallis test ($p<0.001$).

\section*{Text B: Temporal stability of the entropy}
To test the robustness of our results with respect to the time period under consideration, we measured the temporal stability of users' entropy $S$ with a $k-$fold approach.
To this aim, we divided the time window of our study into $k=10$ folds. 
For each user, we computed the entropy averaged over the $k-1$ folds obtained by excluding fold $k$. 
Thus, we obtain a vector $V_1$ containing the entropy of each user averaged over $k-1$ folds. 
Then, we create a second vector $V_2$ with users' entropy measured on the $k-th$ fold. 
The average Pearson correlation between $V_1$ and $V_2$, for all users, is $r = 0.81\pm0.01$ indicating that entropy is a stable metric over time.

\section*{Text C: Sensitivity to the spatial resolution}
In our study, the highest spatial resolution at which we can compute users' mobility metrics is the level of cells (towers clusters). 
To test how our results are impacted by the spatial resolution considered, we computed the semi-partial correlation between the GSE ratio (as defined in the Materials and Methods) and the gender ratio of the mobility metrics (the entropy, $S$, and the set of locations accounting for $80\%$ of a user's activity $\hat{N}_l$) at the cell level. 
The correlation between socio-economic variables and the gender mobility gap, observed for the \textit{comunas}, is still present at the level of cells.
Indeed the Pearson correlation between the GSE ratio and $R_S$ is $r=-0.45$ ($p < 10^{-25}$) and the correlation between the GSE ratio and $R_{\hat{N}_l}$ is $r=-0.40$ ($p <10^{-25}$).

\section*{Text D: Robustness with respect to call activity}
In our study, we selected our users' sample by considering only those who made or received at least one call per day, on average, over the three month study period.
To test the sensitivity of our results to the chosen activity threshold, we also selected a smaller sample of users who made or received at least $3$ calls per day, on average, during the study period, resulting in a total of $254,586$ users.
Overall, the differences in mobility between men and women are also observed in the restricted users' sample: women display a lower mobility entropy and visit a smaller number of locations. 
On average, women visit less distinct locations than men.
Considering the complete set of locations visited by a user, we find women visit about $12$ locations ($95\%$ CI $[11.57, 12.01]$) less than men. 
If we only look at distinct locations that characterize the core of users' daily activity, ($\hat{N}_l$), the difference between men and women becomes $\Delta \hat{N}_l = 2.73$ $[2.68 - 2.79]$.
The average radius of gyration of women, $\langle r_{G} \rangle_F $, is $1.14$ Km ($95\%$ CI $[1.10- 1.17]$) shorter than $\langle r_{G} \rangle_M $.
Also, women's mobility patterns are characterized by a smaller Shannon entropy compared to men, $\Delta S = 0.30$ (95\% CI $[0.29 - 0.30]$).

\section*{Text E: Sensitivity analysis of gender differences in visit patterns}
Here, we discuss several aspects of the analysis of gender differences in visit patterns.
In particular, we describe the sensitivity analysis done to check the robustness of our results and some limitations related to it.

First, the use of CDR data, and the spatial resolution at which we work, only allow us to speak of proximity to a given POI, hence our use of POI densities, estimated via kernel methods, over spatial
scales of ~$1$km or more. We make no claims about gendered visits to places or services that cannot be properly described at a finer spatial scale.

Second, the observed gender differences might be ascribed to gender biases in the \textit{residence locations} of the individuals under study. To rule this out, we repeat the POI analysis by using, for each user, a single location corresponding to the inferred home tower, and we find (Fig.~\ref{SI_POI_home}) that the gender imbalance vanishes for malls and hospitals, and is drastically reduced for taxi stands (we remind that the inferred home location for each user was excluded in the POI analysis reported in Figure $4$ of the main text).
To further investigate this point, we also repeat the POI analysis of Figure $4$ by restricting it to \textit{distant locations} for each user: that is, for each user we 
remove all visited locations within $5$km of the inferred home location. The results (Fig.~\ref{SI_POI_distant}) show that visits to distant places near hospitals and malls are still strongly gendered (in fact, the effect grows slightly stronger), while  the gender imbalance in visits associated to taxi POIs is reduced, as expected if we assume that females use taxi stands in the vicinity of their home more than males.

Third, the observed gender imbalance of Figure 4 might be ascribed to gender differences in \textit{employment} at, e.g., malls and hospitals. To investigate this, we check for the robustness of the POI analysis on removal of the most-frequently-visited non-home location for each user, which we will assume indicates the likely work location of employed individuals. On doing so (Fig.~\ref{POI_work}), we observe no significant change in our results. This robustness holds on removing the top-$k$ most-frequently-visited non-home locations for each user, for several values of $k>1$ (not shown), confirming that the gender differences we report are not associated to users' frequently visited locations (e.g., work place or other relevant place), but rather correspond to places in the long tail of visited locations.

Finally, limitations in the quality of the POI data we use, especially the OpenStreetMap POIs, might in principle influence our results. As far as the \text{mall} and \textit{hospital} POIs are concerned, data quality is actually not an issue: the list of \textit{mall} POIs is generated by looking up towers within the boundary of malls, and the location of malls has been manually verified. The OpenStreetMap \textit{hospital} POIs were successfully checked against official databases of healthcare facilities in Santiago.
It is also important to remark that, for both malls and hospitals, the spatial extent of the facilities ($\sim 1$km$^2$) is comparable to the spatial resolution of the CDR data, hence a CDR record associated to mall or hospital cell tower corresponds with high probability to a user actually visiting those facilities.
As far as \textit{taxi} POIs are concerned, it is challenging to assess spatial biases in their reporting in OpenStreetMap. However, we observe similar gender differences for other type of non-public-transport POIs, such as the \textit{collectivos} shared taxis data (see Fig.~\ref{POI_colectivos}).

\end{document}